\documentclass[aps, pre, superscriptaddress, amsmath,amssymb, reprint, showkeys, ]{revtex4-2}	
	\usepackage{algorithm}
	\usepackage{algpseudocode}
	\usepackage{array}
	\usepackage{mathtools}
	\usepackage[dvipsnames]{xcolor}
	\usepackage{lipsum}
	\usepackage{tikz}
	\usepackage{wrapfig}
	\usepackage[T1]{fontenc}
	\usepackage[utf8]{inputenc}
	\usepackage{amsmath}
	\usepackage{amssymb}
	\usepackage{graphicx}
	\usepackage{refstyle}
	\usepackage{multirow}
	\usepackage{blkarray,booktabs,bigstrut}
	\usepackage{placeins}

	\usepackage[
	citecolor=blue,
	colorlinks,
	linkcolor=blue,
	urlcolor=blue,
	]{hyperref}
	\usepackage{makecell}
	\usepackage{xcolor}
	\usepackage{dcolumn}
	\usepackage{bm}
	\usepackage{tabularx,booktabs}
	\usepackage{multirow}
	\usepackage{float}
	\usepackage[normalem]{ulem}
	\usepackage{multirow}
	\usepackage{blkarray}
	\usepackage{comment}

\begin{document}
	
	
	\title{Overcoming unfairness via repeated interactions in mini-ultimatum game}
	
	\author{Prosanta Mandal}
	\email{prosantam21@iitk.ac.in}
	\address{
		Department of Physics,
		Indian Institute of Technology Kanpur,
		Uttar Pradesh 208016, India
	}
	\author{Arunava Patra}
	\email{arunava20@iitk.ac.in (corresponding author)}
	\address{
		Department of Physics,
		Indian Institute of Technology Kanpur,
		Uttar Pradesh 208016, India
	}

	\author{Sagar Chakraborty}
	\email{sagarc@iitk.ac.in}
	\address{
		Department of Physics,
		Indian Institute of Technology Kanpur,
		Uttar Pradesh 208016, India
	}

\begin{abstract}
	Repeated interactions are ubiquitous and known to promote social behaviour. While research often focuses on cooperation in the Prisoner's Dilemma, experimental evidence suggests repeated interactions also foster fairness. This study addresses a gap in the literature by theoretically modelling the evolution of fairness within a repeated mini-ultimatum game. Specifically, we construct a repeated-game framework where offerers and accepters interact using reactive strategies. We then investigate whether fair reactive strategy pairs are resilient against unfair mutants in a two-species population. By analysing short-term evolutionary stability via the concept of two-species evolutionary stable strategy, we identify a critical effective game length: below this value, fairness is promoted by offerers and accepters who comply with their partner's past actions. Above this critical value, fairness is maintained by `complier' offerers and fair accepters. We also show that specific reactive strategies effectively facilitate the emergence and sustenance of fairness in long-term mutation-selection dynamics. To this end, we develop a two-population stochastic dynamics model---a generalization of classical adaptive dynamics---that accounts for finite population sizes and non-local mutants in the reactive strategy space. 
\end{abstract}

\maketitle

\section{Introduction}
Selfish individuals would share a desirable resource among themselves in ways that maximize their individual gains, thereby leading to obvious situations where unfairness evolves in a population of self-interested individuals. However, it is well documented by a substantial body of experimental and empirical research on both humans and non-human species that fair treatment is not confined to humans, and that animals also exhibit fair behaviour~\cite{M_2010_ZT, BW_2003_NAT, BSW_2005_PRSB, RHVH_2009_PNAS, B_2004_BP, H_1999_BOOK, B_2006_SJR}.  Experiments on chimpanzees, brown capuchin monkeys, and domestic dogs also highlight the presence of fair behaviour in animals~\cite{BW_2003_NAT, BSW_2005_PRSB, RHVH_2009_PNAS}. Furthermore, empirical evidences document a sense of fairness in coyotes and ravens~\cite{B_2004_BP, H_1999_BOOK, B_2006_SJR}. 

Explaining how fairness emerges constitutes an important area of research across disciplines ranging from the social sciences to the biological sciences~\cite{FS_1999_QJE, FF_2003_NAT, AB_2011_JTB}. Understanding its evolutionary roots can contribute to the development of social justice~\cite{B_2005_NJ}, equitable treatment~\cite{BW_2014_Sc}, and fair resource allocation among individuals~\cite{DBA_2017_PO}. In this regard, evolutionary game theory offers a powerful framework for studying how collective outcomes emerge from individual-level interactions~\cite{SP_1973_NAT, M_1974_JTB, S_1988_BOOK}.

A well-known paradigm for modelling fair behaviour within this framework is the ultimatum game (UG)~\cite{NPS_2000_Sc, YLW_2015_EL, HMB_2006_Sc, HCS_2017_PNAS, DBA_2016_EHB}.
In the classical UG, two players must divide a fixed desirable (say, money or food), normalized to one unit, which is initially allocated to one of them, referred to as the offerer. The offerer offers a portion of this amount to the other player, the accepter, who then either accepts or rejects the offer. Acceptance results in the proposed division being implemented, whereas rejection leaves both players with nothing. In principle, the offerer may offer any amount between 0 and 1, and the accepter may ask for any minimum amount within the same range for acceptance. A rational offerer should offer the smallest possible positive amount, while a rational accepter should accept any non-zero offer, since receiving something is preferable to receiving nothing. This outcome constitutes the subgame-perfect Nash equilibrium under the assumption of common knowledge of the game components~\cite{HBB_2005_BBSc}.

Although the UG has received considerable attention in both theoretical and experimental economics since its introduction~\cite{KS_1994_JITE, C_1996_JEBO, GMA_2012_FN, GK_2014_JEBO}, it may be equally relevant for describing biologically realistic scenarios such as food sharing in cooperative hunting and alliance formation within groups. For instance, individuals may negotiate the division of rewards before engaging in joint hunting. In such a case, one dominant individual within the group may determine the division of rewards, while the others may then choose to accept or reject it.

In recent years, considerable research has been devoted to identifying the critical factors that promote fairness in the UG. Reputation-based models demonstrate that reputation updating helps individuals select their interaction partners and prevents proposers from making selfish offers, thereby promoting fair behaviour~\cite{NPS_2000_Sc, AB_2011_JTB, C_2008_RS, ZYC_2023_CSF}. Using the methods of statistical physics, it has been shown that local interactions, network reciprocity, and the spatial organization of strategies can substantially affect the emergence and persistence of fair behaviour~\cite{SPS_2012_PRL, SPS_2012_EPL, DTZ_2011_PHYSA, SBK_2021_PHYSA, PNS_2000_PRSB, M_2007_BOOK, KS_2001_PRSB}.
In addition, unfairness can be mitigated by randomly alternating the roles of offerer and accepter~\cite{WFZW_2013_SR}. Other notable factors that have been shown to foster fairness include variation in stake size~\cite{WCW_2014_SR, ZCL_2018_AMC}, learning~\cite{GBS_1995_GEB}, rare mutation processes, and finite population size~\cite{RTON_2013_PNAS}. Physiological factors such as empathy and emotion have also been studied in the context of the UG and have been found to facilitate the evolution of fairness~\cite{P_2002_BMB, TS_2016_PO}.

Organisms often encounter the same partners multiple times over their lifetime. However, to the best of our knowledge, almost all prior models of the UG have focused on one time interactions and have therefore overlooked the role of repeated encounters between offerers and accepters. Such repeated interactions can take different forms. An offerer may face the same responder over several rounds and accumulate rewards over the course of these interactions, creating an opportunity for the accepter to reject low offers initially in anticipation of more favourable offers later. Alternatively, the roles of offerer and accepter may alternate across rounds, allowing both individuals to participate in determining the division of the resource.

An experiment on the UG in which players interact repeatedly with the same opponent shows that repeated interaction can promote fairness in the population~\cite{S_1999_Experimental}. Since the conflict between the offerer and the accepter becomes more pronounced under repeated interaction, it may push the offerer to behave more fairly. Note that repeated-game frameworks are well established in the literature, but they have primarily focused on the prisoner’s dilemma as the underlying game for explaining the evolution of cooperation~\cite{AH_1981_AAAS, NS_1990_SPRINGER}. A repeated-game framework in the context of the UG remains relatively unexplored, even though such a framework may be particularly useful for investigating history-dependent decision-making and learning through repeated observation in the evolution of fairness.

Therefore, we attempt to establish a repeated UG framework in which players interact with the same partner multiple times, where the offerer responds only to the accepter’s most recent demand and the accepter likewise makes demands solely on the basis of the last offer. All this happens under the shadow of the future~\cite{AH_1981_Sc} that determines the probability that a subsequent round will occur. We then ask whether a fair strategy pair can resist invasion by an unfair mutant pair in a two-species population of offerers and accepters. In this context, we seek to understand how the shadow of the future influences the evolutionary stability of fair strategy pairs.

We investigate this question under two different evolutionary processes. Short-term evolution, which typically refers to the replication-selection process, governs changes in trait frequencies for a fixed set of traits in the population~\cite{SO_2015_DGA, TNP_2006_PRE, N_2006_BOOK}. However, even if a fair strategy pair is evolutionarily stable in the short term, its emergence in the long-term evolutionary process is not guaranteed. We, therefore, also investigate long-term evolution, which is typically a mutation-selection process~\cite{FI_2006_JET, IN_2009_PRSB}. Under this process, a rare mutation appears at each time step in the population and either fixates, thereby resulting in a new monomorphic state, or goes extinct. In this way, the process drives the evolution across different monomorphic population states. The fixation probability of a rare mutant is determined by the replication-selection process. The transient phase from the arrival of a rare mutant to its eventual fixation or extinction is referred to as short-term evolution.

We present the framework of the two-player repeated mini-UG in Section~\ref{sec:repeated 2-player mini-UG}. Subsequently, we present the short-term evolutionary stability of fair strategy pairs against mutants with unfair strategy pairs in Section~\ref{sec:Short-term_evolution}. The framework of long-term evolution in a two-species population and its analysis are presented in Section~\ref{sec:Long-term_evolution}. Our results show that fair reactive strategy pairs resist invasion by unfair mutant pairs, and that a critical value of the shadow of the future determines the performance of fair strategies in both the short and long term. The long-term fairness rate remains high below this critical value. We discuss this contextually in Section~\ref{sec:discussion} to conclude the paper.

\section{Repeated 2-player mini-UG}\label{sec:repeated 2-player mini-UG}
\begin{figure*}
	\centering
	\includegraphics[scale=1.0]{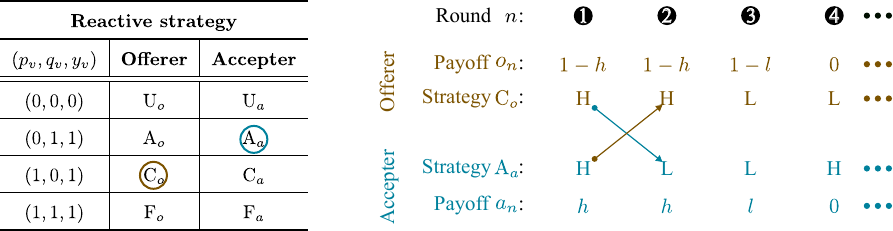}
	\caption{Schematic diagram illustrating the repeated interaction between an offerer and an accepter employing reactive strategies. The pure reactive strategies of the offerer and the accepter are tabulated where $(p_v,q_v,y_v)$ represents the reactive strategy of the offerer and the accepter if the corresponding subscript $v$ is taken to be $o$ and $a$, respectively. The offerer and the accepter are shown to use the reactive strategies $\text{C}_o$ and $\text{A}_a$, respectively. Consequently, in the first round, the offerer makes a high offer (H) and the accepter chooses a high demand (H). Therefore, the offerer and the accepter receive payoffs $o_1=1-h$ and $a_1=h$, respectively, in the first round. In the second round, the offerer complies with the accepter’s previous demand and chooses H, while the accepter anti-complies with the offerer’s previous offer and chooses L. In subsequent rounds, the offerer continues to comply with the accepter’s immediate past demand, whereas the accepter anti-complies with the offerer’s immediate past offer.}
	\label{fig:schematic_reactive}
\end{figure*}
We consider that two players repeatedly interact with each other to distribute a desirable and the repeated interaction occurs in a simultaneous fashion, where both the offerer and the accepter act simultaneously. Furthermore, we assume that both the offerer and the accepter adopt reactive strategies---acting based on the opponent's last move---when playing the repeated game. Since the number of possible offers and demands---regarded as the actions of the offerer and the accepter, respectively---is uncountably infinite, defining reactive strategies would require uncountably many parameters. Therefore, to simplify the analysis without losing non-trivial essence of the game, we focus on a simplified version of the UG, known as the mini-ultimatum game~\cite{GBS_1995_GEB}, and consider it as the underlying game of the repeated interaction.

In the mini-UG, the offerer can offer either high or low, and the accepter also has two available actions: demanding either high or low. We represent the actions high and low of the offerer (or accepter) by H and L, respectively, while the corresponding amounts of offer (or demand) are denoted by $h$ and $l$. This distinction separates the notation for actions from that for the amounts of offers or demands. It is usually considered that the high amount of offer or demand, $h$, corresponds to the equal split $0.5$, since offering more or less than $0.5$ is unfair. The same is true for the accepter: demanding more or less than an equal amount is unfair. The low amount $l$ corresponds to a very small fraction of the desirable amount, close to $0$, and thus $l$ and $h$ satisfy the condition $0<l<h=0.5$.

The payoffs of both the offerer and the accepter are determined by the parameters $l$ and $h$. When the offerer adopts action H and the accepter takes action L, an agreement is successfully reached so that the offerer receives the payoff $1-h$ while the accepter receives the payoff $h$. In contrast, when the offerer adopts action L and the accepter chooses action H, no deal occurs leading to both players receiving zero payoff. This action pair of the offerer and the accepter, (L, H), is referred to as \emph{spiteful} behaviour. If both the offerer and the accepter choose action H, then the offerer receives the payoff $1-h$ while the accepter receives $h$, as the deal takes place. The pair (H, H) corresponds to \emph{fair} behaviour, since the offerer and the accepter are both fair in their offer and demand, respectively. The other possible choice for both the offerer and the accepter is L. In this case, the offerer receives $1-l$ while the accepter receives the remaining amount $l$. This pair of actions, viz., (L,L), leads to an unequal split between them and is therefore associated with \emph{unfair} behaviour. Thus, the strategic interaction between the offerer and the accepter can be modelled through the following bi-matrix game:
\begin{equation}	
	\begin{array}{c|cc}
		& \multicolumn{2}{c}{\text{Accepter}} \\[2pt]
		\text{Offerer} & \text{H} & \text{L} \\ \hline
		\text{H}   & 1-h,\,h & 1-h,\,h \\
		\text{L}   & 0,\,0   & 1-l,\,l  
	\end{array}~.
	\label{eq:UG}
\end{equation}

\subsection{Reactive strategy in mini-UG}
Since the players interact repeatedly by employing reactive strategies, we introduce the notion of reactive strategy in the context of the mini-UG. A reactive strategy is characterized by two conditional probability parameters and one unconditional probability parameter. We denote the reactive strategy of the offerer as $S_o=(p_o, q_o, y_o)$, where $p_o$ and $q_o$ represent the conditional probabilities of playing the action H given that the accepter played H and L in the most recent round, respectively. The parameter $y_o$ represents the unconditional probability of choosing the high offer H in the initial round. Likewise, we denote the reactive strategy of the accepter as $S_a=(p_a, q_a, y_a)$, where $p_a$ and $q_a$ represent the conditional probabilities of playing the action H given that the offerer played H and L in the previous round, respectively. The parameter $y_a$ denotes the unconditional probability of choosing the high demand H in the initial round.

The strategies are called pure reactive strategies if $p_v, q_v \in \{0,1\}$, where $v \in \{o,a\}$. Note that $p_o$ and $q_o$ constitute the reactive strategy space for the offerer, while $p_a$ and $q_a$ form the reactive strategy space for the accepter. The corners of the reactive strategy space correspond to the pure reactive strategies. By adopting the pure reactive strategy $(0,0)$, an offerer (or accepter) chooses the action L in every round. We term this the \emph{unfair strategy} and denote it by $\text{U}_o$ and $\text{U}_a$ for the offerer and the accepter, respectively. In contrast, an offerer (or accepter) chooses the action H in every round when the strategy is $(1,1)$. We call this the \emph{fair strategy} and denote it by $\text{F}_o$ and $\text{F}_a$ for the offerer and the accepter, respectively. In addition to these, an offerer using the strategy $(1,0)$ complies with the accepter's immediate past demand, while an accepter adopting the strategy $(1,0)$ complies with the offerer's immediate past offer. Therefore, it is reasonable to term this strategy the \emph{complier strategy}, and we denote the complier strategies of the offerer and the accepter by $\text{C}_o$ and $\text{C}_a$, respectively. In a similar fashion, we also define an \emph{anti-complier strategy} represented by $(0,1)$. Under this strategy, an offerer defies the accepter and an accepter defies the offerer in every round. We denote this strategy by $\text{A}_o$ and $\text{A}_a$ for the offerer and the accepter, respectively. 

As far as the actions in the initial round required for a complete description of these pure strategies are concerned, they need to be fixed. Clearly, the fair offerer and the fair accepter start with H, while the unfair offerer and the unfair accepter start with the action L. We assume that both the complier and anti-complier strategies play H in the initial round. In summary, we tabulate the pure reactive strategies of the offerer and the accepter and illustrate a repeated interaction between the offerer and the accepter in Fig.~\ref{fig:schematic_reactive}.

\subsection{Payoff }
While the offerer and the accepter interact repeatedly, they receive an average payoff at the end of the repeated interaction. However, realistically, the players are always under the shadow of the future, i.e., the anticipation of future interactions drives current cooperation. How long the ``shadow'' is depends on how high the probability of future interaction is. This notion is traditionally incorporated by considering repeated game with \emph{discounting}: the payoffs are discounted by a discount factor $\delta\in(0,1)$ interpreted as the probability that the next round occurs. For example, the probability that the $n$-th round occurs is given by $\delta^{n-1}$.

At the end of the repeated interaction, the offerer and the accepter receive accumulated payoffs, which are averaged over the expected number of rounds (also known as the effective game length) to determine their average payoffs. The expected number of rounds in a repeated game with discount factor $\delta$ is $\sum_{n=1}^{\infty}\delta^{n-1}=\frac{1}{1-\delta}$. Thus, the expected payoffs of the offerer and the accepter are, respectively, given by
\begin{subequations}
	\begin{eqnarray}
		\pi_o({S_o}, S_a)=(1-\delta)\sum_{n=1}^{\infty} o_{n}\delta^{n-1},
		\label{eq:Average_cum_payoff_offerer}\\
		\pi_a({S_o}, S_a)=(1-\delta)\sum_{n=1}^{\infty} a_n\delta^{n-1},
		\label{eq:Average_cum_payoff_accepter}
	\end{eqnarray}
\end{subequations}
where $o_n$ and $a_n$ are the payoffs of the offerer and the accepter in the $n$-th round, given by $o_n \in \{1-h,0,1-l\}$ and $a_n \in \{h,0,l\}$. The analytical derivation of these expected payoffs is provided in Appendix~\ref{appendix:payoff_fairness_rate}.

\section{Short-term evolution}
\label{sec:Short-term_evolution}
We now proceed to address the evolution of fairness in the regime of short-term evolution. Specifically, we investigate whether, in a heterogeneous population consisting of a fair offerer and accepter subpopulations, an infinitesimal fraction of unfair mutants can outperform the fair heterogeneous population.

\subsection{Set-up}
\label{sec:Short-term_evolution_setup}

We consider a well-mixed heterogeneous population consisting of two infinitely large subpopulations characterized by offerers and accepters. The offerer subpopulation consists of reactive fair and unfair offerers, while the accepter subpopulation consists of reactive fair and unfair accepters. Thus, a reactive fair or unfair offerer can interact with a reactive fair or unfair accepter in the accepter subpopulation, whereas a reactive fair or unfair accepter can interact with a reactive fair or unfair offerer in the offerer subpopulation. Hence, the strategic interaction between fair reactive residents and unfair reactive mutants can be mathematically represented by a bi-matrix game.

Suppose the resident population state is given by $(S_o, S_a)$ and the mutant state by $(\tilde{S}_o, \tilde{S}_a)$. Then, the strategic interaction between the offerer and the accepter can be described by the following repeated-game payoff matrices:
\begin{subequations}
	\begin{align}
		{\sf O} &=
		\begin{bmatrix}
			\pi_o(S_o , S_a) & \pi_o(S_o , \tilde{S}_a)\\
			\pi_o(\tilde{S}_o , S_a) & \pi_o(\tilde{S}_o , \tilde{S}_a)
		\end{bmatrix} \\
		\text{and}\qquad
		{\sf A} &=
		\begin{bmatrix}
			\pi_a(S_o, S_a) & \pi_a(\tilde{S}_o, S_a)\\
			\pi_a(S_o, \tilde{S}_a) & \pi_a(\tilde{S}_o, \tilde{S}_a)
		\end{bmatrix},
	\end{align}
	\label{eq:Repeated_game_matrix}
\end{subequations}%
where the payoff elements are obtained from Eqs.~(\ref{eq:Average_cum_payoff_offerer}) and (\ref{eq:Average_cum_payoff_accepter}).

It is worth noting that the pairs of reactive strategies of the offerer and the accepter leading to fairness in the population are $(\text{F}_o, \text{F}_a)$, $(\text{F}_o, \text{C}_a)$, $(\text{C}_o, \text{F}_a)$, and $(\text{C}_o, \text{C}_a)$. This is because adopting any of these strategies leads to the action pair (H, H) being played in every round of the repeated interaction. In contrast, the pairs of reactive strategies that lead to unfairness in the population are $(\text{U}_o, \text{U}_a)$, $(\text{U}_o, \text{C}_a)$ and $(\text{C}_o, \text{U}_a)$, since these strategy pairs result in the action pair (L, L) being played in every round (except first round in case of the latter two strategy pairs) of the repeated game.

For a single homogeneous population playing a symmetric game, the classical concept of an evolutionarily stable strategy (ESS) was introduced by Maynard Smith and Price~\cite{SP_1973_NAT}. Defining ESS in asymmetric games requires a different treatment because distinct populations may occupy different roles and receive payoffs from different payoff matrices~\cite{C_1992_BOOK, C_1996_TPB, CT_2014_PNAS, DCPC_2025_Chaos}. Our model belongs to this class: offerers and acceptors form two distinct subpopulations that interact through the payoff matrices ${\sf O}$ and ${\sf A}$, respectively. The evolutionary interaction is, therefore, a two-subpopulation bimatrix game rather than a single-population symmetric game. For such frequency-dependent two-population interactions, Cressman~\cite{C_1996_TPB} proposed a local stability criterion requiring that, for every nearby population state, at least one of the two populations has a fitness advantage in the direction of the resident state. It should be pointed out that several related notions of ESS, including so-called \textit{and}-, \textit{sum}-, and \textit{or}-type conditions to define it, have been considered for asymmetric games---they have been extensively compared and contrasted in a recent review~\cite{DCPC_2025_Chaos}. Here, we employ the \textit{or}-type condition (in other words, Cressman's definition), which is appropriate for the coupled fitness responses of our two interacting subpopulations where mutants may appear either in one or both subpopulations; in contrast, the more restrictive \textit{and}-type condition (first introduced by Selten~\cite{selten1980note}) requires the corresponding fitness inequality to hold simultaneously in both subpopulations with mutants not allowed to appear in only one subpopulation. It is important to note that both definitions are equivalent to the extent that neither allow mixed strategies to be ESS.

Let $\boldsymbol{x}=\{x_{S_o}, x_{\tilde{S}_o}\}$ represent the strategy of an offerer in the offerer subpopulation, where $x_{S_o}$ and $x_{\tilde{S}o}$ are the probabilities of adopting the types $S_o$ and $\tilde{S}_o$, respectively. Likewise, $\boldsymbol{y}=\{y_{S_a}, y_{\tilde{S}_a}\}$ denotes the strategy of an accepter in the accepter subpopulation, where $y_{S_a}$ and $y_{\tilde{S}_a}$ are the probabilities of being of types $S_a$ and $\tilde{S}_a$, respectively. Then, $(\boldsymbol{x}, \boldsymbol{y})$ refers to an arbitrary (mixed) strategy pair in the two-species population. A strategy pair $(\boldsymbol{\hat{x}}, \boldsymbol{\hat{y}})$ is an ESS if and only if 
\begin{equation}
	\text{either}~~~\boldsymbol{\hat{x}}\cdot{\sf O}\boldsymbol{y}> 	\boldsymbol{x}\cdot{\sf O}\boldsymbol{y}~~~\text{or}~~~\boldsymbol{\hat{y}}\cdot{\sf A}\boldsymbol{x}> \boldsymbol{y}\cdot{\sf A}\boldsymbol{x}
	\label{eq:ess_cond}
\end{equation} 
for every strategy pair $(\boldsymbol{x}, \boldsymbol{y})$ that is sufficiently close to, but not equal to $(\boldsymbol{\hat{x}}, \boldsymbol{\hat{y}})$.

This definition of ESS is general and not confined to reactive strategies. In principle, one could this ask whether a strategy is ESS in any bimatrix game. For example, one can check that the unfairness (L, L) is a ESS whereas the fairness (H, H) is not a ESS in the underlying game  Eq.~(\ref{eq:UG}). This implies that, in the absence of any repeated interaction, unfairness should evolve---a fact which is obviously inconsistent with empirically observed fairness all around us. Thus, as a central theme of this paper we are curious whether repeated interactions can overcome unfairness by enabling fair reactive strategies to become ESS against unfair mutants.

It may be pointed out that the implicit replication-selection process in infinite well-mixed population is traditionally modelled as replicator dynamics~\cite{TJ_1978_MB, SSH_1981_BC}. But since ESS is proven~\cite{CT_2014_PNAS, DCPC_2025_Chaos} to be  asymptotically stable state of bimatrix game under this short-term dynamics, we need not explicitly deal with the dynamics; direct knowledge of the ESS is enough to conclude what the dynamics would anyway give in the long run.

\subsection{Results}\label{sect:short_term_results}
\begin{figure}[h]
	\centering
	\includegraphics[scale=0.49]{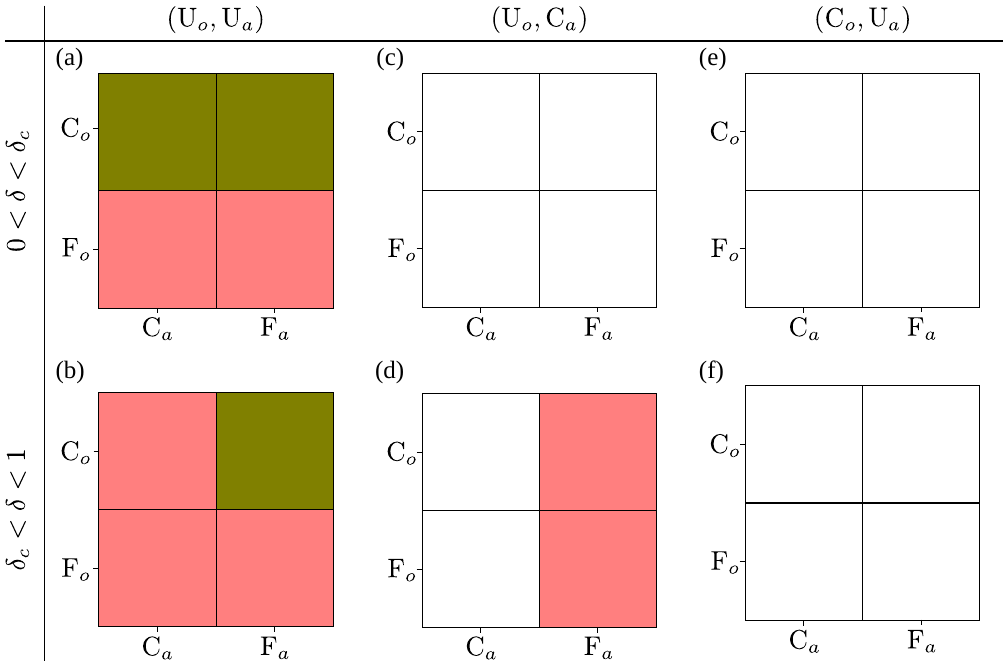}
	\caption{Evolutionary stability of fairness in repeated mini-UG: Here we showcase ESS phase diagrams of fair strategy pairs [$(\text{F}_o, \text{F}_a)$, $(\text{F}_o, \text{C}_a)$, $(\text{C}_o, \text{F}_a)$, and $(\text{C}_o, \text{C}_a)$] versus unfair strategy pairs [$(\text{U}_o, \text{U}_a)$, $(\text{U}_o, \text{C}_a)$, and $(\text{C}_o, \text{U}_a)$] in an infinite two-species population. {Subplots (a)--(b), (c)--(d), and (e)--(f) represent the evolutionary stability of fair strategies against the unfair strategy pairs $(\text{U}_o,\text{U}_a)$, $(\text{U}_o,\text{C}_a)$, and $(\text{C}_o,\text{U}_a)$, respectively.}  The upper and lower panels correspond to the discount factor ranges $0 < \delta < \delta_c$ and $\delta_c < \delta < 1$, respectively. The olive colour indicates that both the fair and unfair strategies are ESSes against each other, whereas the light red colour indicates that only the unfair strategy is ESS. The no colour (or white) boxes represent the absence of any ESS.}
	\label{fig:ESS_short_term}
\end{figure}

We present our results using ESS phase diagrams, which illustrate the outcomes of the evolutionary stability analysis via distinct colour codes. To construct the ESS phase diagram, we separately derive a bi-matrix game for each combination of fair and unfair repeated-game strategies and then determine the ESS from the corresponding payoff matrices using the condition in Eq.~(\ref{eq:ess_cond}). Since there are four fair strategy pairs and three unfair strategy pairs, we construct a total of twelve bi-matrix games to obtain the ESS phase diagrams shown in Fig.~\ref{fig:ESS_short_term}.

We find that fair reactive strategies can prevent invasion by unfair reactive mutants. However, the ability of fair strategy pairs to resist invasion of unfair mutants depends on the discount factor. In particular, we identify a critical discount factor, $\delta_c$, which is determined by the values of high offer (or demand) and low offer (or demand). The critical value is given by $\delta_c = \frac{1 - h}{1 - l}$. Below the critical discount factor $\delta_c$, two fair strategy pairs $(\text{C}_o, \text{C}_a)$ and  $(\text{C}_o, \text{F}_a)$ are ESSes against the unfair strategy  pair $(\text{U}_o, \text{U}_a)$. In contrast, for $1>\delta > \delta_c$, only the fair strategy pair $(\text{C}_o, \text{F}_a)$ remains evolutionarily stable against invasion by $(\text{U}_o, \text{U}_a)$. It can be seen by comparing Fig.~\ref{fig:ESS_short_term}(a) and Fig.~\ref{fig:ESS_short_term}(b). The strategy pair $(\text{C}_o, \text{F}_a)$ is also evolutionarily stable at the critical discount factor $\delta = \delta_c$. It is worth mentioning that, at $\delta = \delta_c$, the ESS phase diagram against $(\text{U}_o, \text{U}_a)$ coincides with Fig.~\ref{fig:ESS_short_term}(b), while the ESS phase diagram against $(\text{U}_o, \text{C}_a)$ matches Fig.~\ref{fig:ESS_short_term}(c).

The performance of fair reactive strategies also depends on the nature of the unfair reactive mutants. A resident two-species population adopting the fair strategy pair $(\text{C}_o, \text{F}_a)$ can resist invasion by the unfair mutant pair $(\text{U}_o, \text{U}_a)$. In contrast, the same resident population cannot prevent invasion by the unfair mutant pair $(\text{U}_o, \text{C}_a)$; compare  Fig.~\ref{fig:ESS_short_term}(b) and Fig.~\ref{fig:ESS_short_term}(d). A similar observation holds for resident population adopting the fair strategy pair $(\text{C}_o, \text{C}_a)$, irrespective of the value of the discount factor. Finally, neither of the fair strategy pairs is ESS against the unfair mutant pair $(\text{C}_o,\text{U}_a)$, regardless of the discount factor---see Fig.~\ref{fig:ESS_short_term}(e) and Fig.~\ref{fig:ESS_short_term}(f).

It is important to note that the unfair strategy pair $(\text{U}_o, \text{C}_a)$ is not a ESS against all possible fair mutants, and the fair mutants are also not ESS against the unfair strategy pair $(\text{U}_o, \text{C}_a)$ for $\delta<\delta_c$ [see Fig.~\ref{fig:ESS_short_term}(c)]. This result also holds at $\delta=\delta_c$. Consequently, no ESS exists in the population when the discount factor lies in the range $0<\delta \leq \delta_c$. However, when the discount factor exceeds the critical value ($\delta > \delta_c$), the unfair strategy pair regains evolutionary stability against the fair mutant pairs $(\text{C}_o, \text{F}_a)$ and $(\text{F}_o, \text{F}_a)$.

In summary, we find two important results: the complier strategy pair $(\text{C}_o, \text{C}_a)$ is evolutionarily stable against $(\text{U}_o, \text{U}_a)$ only when the effective game length is sufficiently short, whereas $(\text{C}_o, \text{F}_a)$ outperforms the mutant pair $(\text{U}_o, \text{U}_a)$ regardless of the effective game length. We next provide an intuitive explanation of these results. 

The evolutionary stability of $(\text{C}_o, \text{C}_a)$ against $(\text{U}_o, \text{U}_a)$ can be understood as follows. The complier accepter begins with a high demand, thereby punishing the unfair offerer while reaching agreement with the complier offerer in the initial round. However, by complying with the unfair offerer in subsequent rounds, it lowers its own rewards and allows the unfair offerer to increase its payoff. As a result, over long repeated interactions, the unfair offerer gradually derives a fitness advantage from the complier accepter’s compliance. Meanwhile, the complier offerer accepts the complier accepter’s high demand from the outset and therefore earns less than the unfair offerer in future rounds, although the complier accepter obtains higher rewards than the unfair accepter against the complier offerer. Consequently, beyond a critical value of the shadow of the future, the fitness of the complier strategy pair is such that it can no longer resist invasion by the unfair strategy pair.

Likewise, the survival of the strategy pair $(\text{C}_o, \text{F}_a)$ against $(\text{U}_o, \text{U}_a)$ regardless of the effective game length can be explained as follows. In this pair, the accepter acts as a tough accepter by always demanding high. When an infinitesimal fraction of unfair mutants appears, the tough accepter punishes the unfair offerer by rejecting the unfair offer in every round, thereby reducing its fitness. At the same time, it reaches an agreement with the complier offerer in every round, enabling the latter to resist invasion by the unfair offerer. In addition, the fair accepter extracts higher rewards from the complier offerer than the unfair accepter does. As a result, the strategy pair $(\text{C}_o, \text{F}_a)$ resists invasion by unfair mutants.

\subsection{Critical discount factor, $\delta_c$}
To illustrate how the critical discount factor $\delta_c$ arises from the stability condition in Eq.~(\ref{eq:ess_cond}), consider the resident strategy pair $(\text{C}_o,\text{C}_a)$ and the mutant pair $(\text{U}_o,\text{U}_a)$.
When offerer and acceptor play against each other repeatedly ad infinitum, the repeated-game expected payoffs they get can be conveniently represented as a bi-matrix:
\begin{equation}
	\begin{array}{c|cc}
		& \text{C}_a & \text{U}_a \\ \hline
		\text{C}_o
		& 1-h,\,h
		& (1-h)+(h-l)\delta,\,
		h+(l-h)\delta \\[1mm]
		\text{U}_o
		& (1-l)\delta,\,l\delta
		& 1-l,\,l
	\end{array},
	\label{eq:Repeated_UG_CU}
\end{equation}
where in each pair of payoffs separated by comma, the first one is for the offerer and the second one is for the acceptor.

Consider a small perturbation of the resident state, $\hat{x}=\hat{y}=(1,0)$:
\begin{equation}
	x=y=(1-\epsilon,\epsilon),\qquad
	0<\epsilon\ll1.
\end{equation}
The payoff differences in the offerer subpopulation  and the accepter subpopulation, respectively, are
\begin{subequations}
\begin{align}
	\hat{x}\cdot{\sf O}y-x\cdot{\sf O}y
	&=
	\epsilon\left[(1-h)-(1-l)\delta\right]
	+\mathcal{O}(\epsilon^2),\label{eq:tresnon}\\
	\hat{y}\cdot{\sf A}x-y\cdot{\sf A}x
	&=
	\epsilon\,\delta(h-l)+\mathcal{O}(\epsilon^2).
\end{align}
\end{subequations}
A close inspection of the above reveals that for condition~(\ref{eq:ess_cond}) to be fulfilled, since $h>l$ and $\delta>0$, the nontrivial stability threshold is solely determined by the offerer subpopulation: defining,
\begin{equation}
	\delta_c=\frac{1-h}{1-l},
\end{equation}
we conclude that for $\delta<\delta_c$, the resident pair $(\text{C}_o,\text{C}_a)$ satisfies the stated two-population evolutionary stability condition against the mutant pair $(\text{U}_o,\text{U}_a)$. The stability conditions for the other resident strategy pairs can be obtained analogously.

\section{Long-term evolution}
\label{sec:Long-term_evolution}
Although we have established that fairness emerges and persists through repeated interactions in the short term under the replication–selection process, this does not ensure the long-term evolution of fairness under the mutation-selection regime. In what follows, we examine whether the reactive strategies that promote the emergence of fairness can persist in the long-term mutation-selection process.

\subsection{Set-up}
To this end, we  consider a finite, well-mixed heterogeneous population consisting of offerer and accepter subpopulations of sizes $N_o$ and $N_a$, respectively. At any time step, all offerers make the same offer and all accepters demand the same amount; thus, each subpopulation is monomorphic. The initial state of the population is taken to be $(\text{U}_o, \text{U}_a)$, with unfair offerers and unfair accepters, since our primary interest lies in the evolution of fairness. The emergence of fairness requires mutations in this unfair population, and evolutionary systems typically undergo rare mutation processes. Accordingly, we introduce a single mutant at each time step. Here, a rare mutation implies that a new mutant does not arise until the fixation or extinction of the previous mutant is complete.

In a heterogeneous population, mutants can appear in either or both subpopulations. If mutants emerge in both subpopulations, this can lead to three possible scenarios: the extinction of mutants from both subpopulations, the fixation of mutants in both subpopulations, or the fixation of a mutant in either subpopulation. The latter two scenarios give rise to three new population states. For instance, if a mutant pair denoted by $(\tilde{S}_{o}, \tilde{S}_{a})$ arises in a resident population state characterized by $(S_{o}, S_{a})$, then it can result in three new population states: $(\tilde{S}_{o}, \tilde{S}_{a})$, $(\tilde{S}_{o}, S_{a})$, and $(S_{o}, \tilde{S}_{a})$. Therefore, mutations in both subpopulations lead to transitions of the resident population state to any of these three population states. If a mutant emerges in only one subpopulation, then it gives rise to only one new population state, as the mutant can either fixate or go extinct in that subpopulation while the other subpopulation remains unchanged. For example, when the mutant pair $(\tilde{S}_{o}, S_{a})$  occurs in the resident population $({S}_{o}, {S}_{a})$, it leads to one new population state $(\tilde{S}_{o}, S_{a})$. Note that, for convenience, we have used the term `mutant pair' even when mutant appears in one subpopulation.

In passing, we note that an existing model of long-term evolutionary dynamics in asymmetric games typically considers mutations arising in only one subpopulation at a time~\cite{VH_2016_JET}. In contrast, we also allow the simultaneous occurrence of mutants in both subpopulations in order to maintain consistency between long-term evolutionary dynamics and the definition of ESS in two-species populations. According to the definition of an ESS, a resident state must be robust against all possible mutant invasions, regardless of whether the mutants arrive in one or both subpopulations. Certainly, the simultaneous occurrence of mutants in both subpopulations is rarer than the occurrence of a mutant in a single subpopulation. This fact is incorporated into our model since the probability of simultaneous mutations is given by the product of the mutation rates in the two subpopulations. The detailed formalism is provided in Appendix~\ref{appendix:long_term_evolution_two_species}.

Here, a mutant in each subpopulation is an individual adopting a new strategy, and mutants arise successively. Once the fixation or extinction of the previous pair of mutants in both subpopulations is complete, another random pair of mutants emerges in the next time step and either takes over the respective subpopulations or goes extinct. In this way, the mutation--selection process recurs indefinitely. During its transient phase, the resident and mutant compete through the replication--selection process, and the strength of selection is quantified by the parameter $w$. It takes values from zero to infinity, and random drift dominates when its value is close to zero. Increasing its value increases the weight of payoff in fitness. Eventually, the selection strength and payoff determine the probability of fixation of the mutant pair. This probability is quantified by the fixation probability, and its formulation employing the replication--selection process is discussed in Appendix~\ref{appendix:birth-death_process}. Note that each subpopulation becomes monomorphic after the fixation or extinction of a mutant is complete, and the two subpopulations transition from one monomorphic state to another on a long evolutionary time scale.

Let us denote the strategy set of the offerer by $\mathcal{S}_o=\{\text{F}_o, \text{A}_o, \text{C}_o, \text{U}_o\}$ and and that of the accepter by $\mathcal{S}_a=\{\text{F}_a, \text{A}_a, \text{C}_a, \text{U}_a\}$, from which mutants for the offerer and the accepter subpopulations are randomly drawn, respectively. The mutation rates in the offerer and the accepter subpopulations are denoted by $\mu_o$ and $\mu_a$, respectively. Clearly, there are 16 possible population states, since a population state is characterized by a pair of strategies of the two subpopulations, $({S}_{o}, {S}_{a})$, where $S_o\in\mathcal{S}_o$ and $S_a\in\mathcal{S}_a$.

Now, let us define the \emph{fairness rate} corresponding to $({S}_{o}, {S}_{a})$ by $\gamma({S}_{o}, {S}_{a})$, which measures the frequency of the action pair (H, H) when an offerer with strategy ${S}_{o}$ and an accepter with strategy ${S}_{a}$ play repeated games. As the population evolves among these 16 states, the fairness rate varies over time. If we let $\alpha_{({S}_{o}, {S}_{a})}^{(t)}$ denote the frequency of the monomorphic population state with strategy pair $({S}_{o}, {S}_{a})$ at time $t$ in the mutation–selection process, then the \emph{average fairness level} in the population at time $t$ is given by
\begin{equation}
	\langle \gamma(t) \rangle = \sum_{\substack{ S_o\in\mathcal{S}_o\\S_a\in\mathcal{S}_a}} \alpha_{({S}_{o}, {S}_{a})}^{(t)}\, \gamma(S_o, S_a).
	\label{eq:avg_fairness_rate}
\end{equation} 
The detailed expression for the fairness rate $\gamma({S}_{o}, {S}_{a})$ is provided in Appendix~\ref{appendix:payoff_fairness_rate}, and the analytical derivation of $\alpha_{({S}_{o}, {S}_{a})}^{(t)}$ from the mutation--selection process in a two-species population is presented in Appendix~\ref{appendix:long_term_evolution_two_species}. 

\subsection{Results}
\begin{figure}
	\centering
	\includegraphics[scale=0.48]{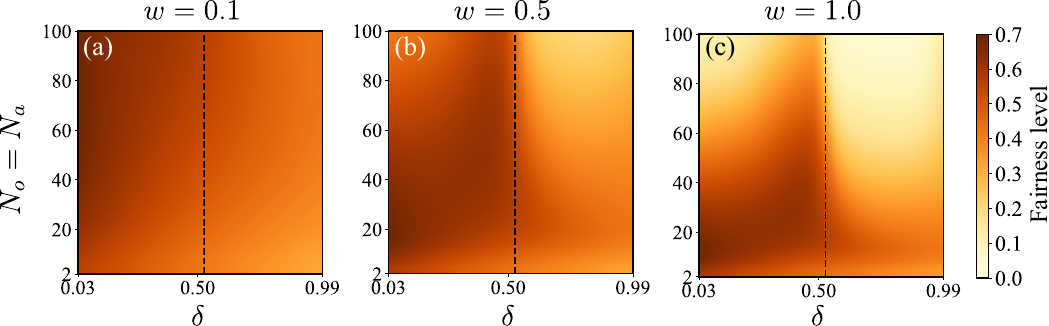}
	\caption{Emergence of fairness via repeated interactions in mutation-selection regime. The long-term fairness level is shown across the parameter space of population size and discount factor for three different selection strengths, $w \in \{0.1, 0.5, 1.0\}$. The population sizes and discount factor are varied within the ranges $N_o = N_a \in [2, 100]$ and $\delta \in [0.03, 0.99]$, respectively. The colour bar represents the average fairness rate in the population. As the colour gradient transitions from light to dark, the level of fairness increases.}	
	\label{fig:fairness_N_df}
\end{figure}

To analyse the long-term evolutionary dynamics, we assume, without loss of generality, that the two subpopulations have equal sizes, thereby reducing the number of parameters. Furthermore, we fix the high offer or demand at $h = 0.5$, as offering or demanding more or less than an equal proportion is considered unfair. The low offer or demand is set to $l = 0.05$. The mutation rates in both subpopulations are assumed to be equal; in particular, we set $\mu_o = \mu_a = 10^{-2}$ for computational efficiency.  Smaller mutation rates only increase the convergence time to the stationary distribution without qualitatively affecting the results.  We then present the long-term fairness level across population sizes $N_o = N_a \in [2, 100]$ and discount factors $\delta \in [0.03, 0.99]$ for three different selection strengths, $w \in \{0.1, 0.5, 1.0\}$, as shown in Fig.~\ref{fig:fairness_N_df}.

We emphasize that the fixation probabilities used in our long-term evolutionary analysis are obtained numerically from the underlying stochastic birth--death process (see Appendix~\ref{appendix:birth-death_process}) and do not rely on weak-selection approximation whose analytical expression~\cite{SO_2015_DGA} for fixation probabilities is valid only when $w\ll 1/(N_o + N_a)$. For example, for $N_o=N_a=100$, this requires $w\ll10^{-2}$. We therefore do not interpret the values $w=0.1$, $0.5$, and $1.0$ as the weak-selection approximation. We however do not include very small selection strengths, such as $w=0.001$, in Fig.~\ref{fig:fairness_N_df}, since in this regime selection becomes weak relative to stochastic drift and the effects of the underlying game on the long-term population distribution become progressively less pronounced. Indeed, the influence of stochastic drift is already appreciable for $w=0.1$, as can be seen in Fig.~\ref{fig:fairness_N_df}. We therefore focus on $w=0.1$, $0.5$, and $1.0$ to illustrate the effects of increasing selection strength beyond the drift-dominated regime.

To exclusively understand the emergence of fairness, we initialize both the offerer and the accepter subpopulations with unfair individuals, i.e., the initial population state is taken to be $(\text{U}_o, \text{U}_a)$, and we let the system evolve. After $10^5$ generations, we record the distribution of population states and compute the average fairness rate using Eq.~(\ref{eq:avg_fairness_rate}). We perform this analysis for each combination of discount factor and population size to generate the plots in Fig.~\ref{fig:fairness_N_df}.

\begin{figure*}
	\centering
	\includegraphics[scale=0.85]{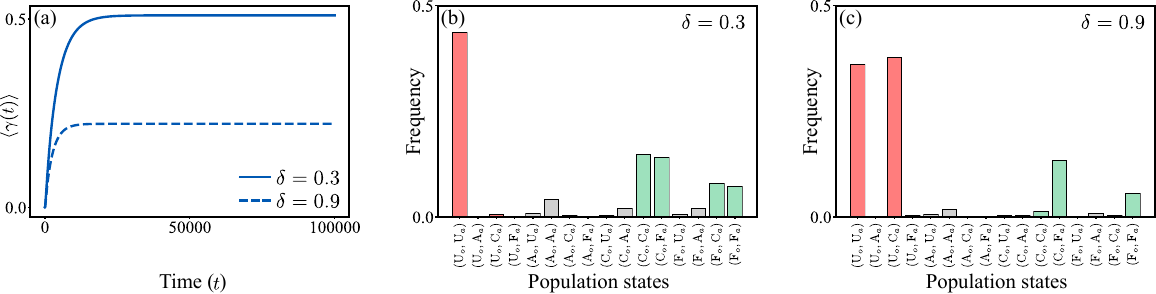}
	\caption{The long-term fairness level increases as discount factor decreases. The fairness level and the frequencies of reactive strategy pairs in long-term are illustrated for two discount factors, $\delta = 0.3$ and $\delta = 0.9$, while keeping the selection strength fixed at $w = 1.0$. Subplot (a) shows the fairness level in the population over time, where the dotted and solid lines represent the fairness levels for $\delta = 0.3$ and $\delta = 0.9$, respectively. Subplots (b) and (c) depict the distribution of strategy pairs after $10^5$ generations. The light red colour denotes the unfair strategy pairs, whereas the sea green colour represents the fair strategy pairs. The grey colour corresponds to the remaining strategy pairs. }	
	\label{fig:freq_station_dist_df}
\end{figure*}%

We observe that, for selection strength $w = 0.1$, the colour gradient remains nearly uniform across population sizes and discount factors. Thus, the fairness level does not depend significantly on either the discount factor or the population size at the relatively low selection strength, since random drift influences individual fitness more strongly than the payoff obtained from game interactions. As the selection strength increases to $w = 0.5$, the fairness level begins to vary across population sizes and discount factors [see Fig.~\ref{fig:fairness_N_df}(b)], as the payoff from game interactions starts to dominate over random drift.

It may be noted that the fairness level decreases as the population size increases, and this holds for both low and high discount factors. Increasing the population size reduces stochastic fluctuations, and the unfair strategy pair $(\text{U}_o, \text{U}_a)$ is always ESS against fair reactive strategy pairs in an infinite population, leading to a decline in the fairness level. The variation in the fairness level becomes more conspicuous as the selection strength increases further [see Fig.~\ref{fig:fairness_N_df}(c)], since selection is largely driven by the payoff from game interactions. For stronger selection ($w = 0.5, 1.0$), we also observe that the fairness level decreases with increasing discount factor. Moreover, for large population sizes, there is a sudden shift in the fairness level beyond a particular value of the discount factor. Notably, this value coincides with the critical discount factor $\delta_c=\frac{1-h}{1-l}$ identified in the short-term evolution.

To better understand this, we select two representative points from Fig.~\ref{fig:fairness_N_df}(c). One point corresponds to a discount factor $\delta = 0.3$, which lies below the critical value $\delta_c$, while the other corresponds to $\delta = 0.9$, which lies above $\delta_c$. Both the points are chosen for the same population size, $N_o = N_a = 50$. For these two parameter sets, we illustrate the evolution of fairness and the corresponding stationary distributions of population states in Fig.~\ref{fig:freq_station_dist_df}. Fig.~\ref{fig:freq_station_dist_df}(a) shows the fairness level over time, whereas Figs.~\ref{fig:freq_station_dist_df}(b) and \ref{fig:freq_station_dist_df}(c) display the distribution of population states after $10^5$ generations for discount factors $\delta=0.3$ and $\delta=0.9$, respectively.

As shown in Fig.~\ref{fig:freq_station_dist_df}(b), the fair strategy pairs $(\text{C}_o, \text{C}_a)$ and $(\text{C}_o, \text{F}_a)$ are predominant in the population when the discount factor is $\delta = 0.3$. In contrast, when the discount factor increases to $\delta = 0.9$, the frequency of the population state $(\text{C}_o, \text{C}_a)$ approaches zero; see Fig.~\ref{fig:freq_station_dist_df}(c). This difference can be understood from the short-term evolutionary stability analysis. When the discount factor lies below the critical value $\delta_c$, the strategy pairs $(\text{C}_o, \text{C}_a)$ and $(\text{C}_o, \text{F}_a)$ are ESS against the unfair mutant pair $(\text{U}_o, \text{U}_a)$ [see Fig.~\ref{fig:ESS_short_term}(a)]. Consequently, a resident population adopting $(\text{C}_o, \text{C}_a)$ or $(\text{C}_o, \text{F}_a)$ hinders the fixation of unfair mutants in the mutation–selection process, allowing both strategies to persist in the long term. In contrast, for the higher discount factor $\delta = 0.9$, $(\text{C}_o, \text{F}_a)$ remains ESS, whereas $(\text{C}_o, \text{C}_a)$ loses its evolutionary stability once the discount factor exceeds $\delta_c$ [see Fig.~\ref{fig:ESS_short_term}(a) and Fig.~\ref{fig:ESS_short_term}(b)]. As a result, a resident population with strategy pair $(\text{C}_o, \text{C}_a)$ is unable to resist the fixation of unfair mutants, thereby reducing its prevalence in the long term and causing it to approach zero. In passing, it is interesting to note that the unfair strategy pair $(\text{U}_o, \text{C}_a)$ emerges in the long-term evolution for a high discount factor $\delta = 0.9$, since it becomes ESS against fair mutants in the short-term evolution when the discount factor exceeds the critical value $\delta_c$ [see Fig.~\ref{fig:ESS_short_term}(c) and Fig.~\ref{fig:ESS_short_term}(d)]. 

Summing up, the long-term fairness level is relatively higher in repeated interactions with a short effective game length, and it is predominantly driven by strategy pairs that include the complier offerer and either a complier accepter or a tough accepter. For repeated games with a high effective game length, the two unfair states become more frequent in the population, which leads to a decline in the overall fairness level [Fig.~\ref{fig:freq_station_dist_df}(a)]. In this case, the complier offerer and the fair accepter predominantly drive the long-term fairness. Thus, the long-term results are consistent with the short-term outcomes.

\section{Discussion}
\label{sec:discussion}
Understanding how simple microscopic interaction rules generate nontrivial macroscopic behaviour is a central theme of statistical physics and complex-systems research~\cite{A_1972_Sc, CFL_2009_RMP}; in this regard, evolutionary game theory provides a natural framework for connecting individual-level strategic interactions with population-level evolutionary outcomes \cite{N_2006_BOOK, NST_2004_NAT}. Our present work provides a microscopic-to-macroscopic description of the emergence of fairness. At the microscopic level, individuals repeatedly interact in pairs, while stochastic birth--death events and mutations govern evolutionary changes in the finite population. These stochastic microscopic processes collectively determine fixation probabilities and, consequently, the stationary distribution of strategy pairs in the long-time limit. Fairness can therefore be interpreted as an emergent population-level outcome generated by the underlying stochastic interactions among individuals.

Some previous studies have considered iterated versions of the UG, in which behavioural decisions may depend on information acquired during previous interactions. In particular, Almeida et al.~\cite{ASM_2014_PHYSA} investigated an iterated UG with both memoryless and one-step-memory strategies, demonstrating that short-term information about previous interactions can influence the evolutionary dynamics of the game. However, a repeated-game framework in the context of the UG remains relatively unexplored, even though such a framework may be particularly useful for investigating history-dependent decision-making and learning through repeated observation in the evolution of fairness. In particular, the role of explicit reactive behaviour---in which the offerer's current offer is conditioned on the accepter's previous demand and the accepter's current demand is conditioned on the offerer's previous offer---has received little attention. Here, we address this issue by introducing role-specific reactive strategies in a repeated UG and investigating whether such direct responses to a partner's preceding action can support the emergence and evolutionary stability of fairness.

With a view to succinctly presenting the essence of the main results, in this paper, we have considered a simpler yet non-trivial version of UG---mini-UG, in which the offerer offers and the accepter demands either a high amount, corresponding to half of the desirable amount, or a low amount, corresponding to a proportion close to zero. This reduction keeps the repeated-game framework simple while preserving the essential features of the problem. It is further assumed that offerers and accepters adopt such reactive strategies in the repeated interaction.

Having identified the reactive strategy pairs capable of sustaining fairness, we have then investigated their short-term evolutionary stability against infinitesimal unfair reactive mutants in a well-mixed infinite two-species population of offerers and accepters. However, such short-term evolutionary stability does not necessarily ensure their persistence in the long term. For completeness, we therefore examined the evolution of fair reactive strategies in the long-term mutation–selection process.

The fair strategy pair in which the offerer complies with the accepter’s demand and the accepter complies with the offerer’s offer is a ESS against unfair reactive mutants and promotes fairness. Notably, we identify a critical discount factor below which the complier strategy is a ESS. This critical value depends on the values of high and low offers or demands. In contrast, the strategy pair consisting of a complier offerer and a tough accepter—who demands a fair proportion—fosters fairness even for high discount factors, as it remains a ESS against unfair reactive mutants irrespective of the discount factor. These two fair reactive strategy pairs also predominantly drive the evolution of fairness in the long term. In particular, for large population sizes, we observe a conspicuous shift in the long-term fairness level around the critical discount factor. In conclusion, repeated interactions with a small effective game length facilitate the emergence of fairness in both short- and long-term evolutions.

An experimental study~\cite{S_1999_Experimental} on repeated interactions in the UG reports that fair behaviour is driven by \emph{tough} accepters who reject many fair offers to build a tough reputation for extracting a large proportion of the share when offerers offer any amount from zero to one. Our analysis reveals similar fact: Both short-term and long-term fairness are predominantly driven by the fair accepter---which may be seen analogous to a tough accepter in our setup---and the complier offerer in long repeated interactions. Notably, the analytical framework developed to study the mutation–selection process in a two-species population may facilitate future investigations of emergent behaviour in heterogeneous populations. Our results are not analogous to, but instead differ subtly from, existing findings on cooperation in repeated-game settings. In particular, while a large effective game length is typically required for reciprocal strategies to become evolutionarily stable and sustain cooperation~\cite{AH_1981_AAAS}, a small effective game length is sufficient for the complier strategy to evolve and foster fairness.

Our modelling framework, while deliberately minimal, captures the essential features of a broad class of repeated dyadic allocation problems and, thus, admits concrete application domains. On the economic side, it directly relates to repeated bilateral bargaining scenarios---such as long-term employer--employee wage negotiations or recurring buyer--seller transactions---where prior offers and demands naturally shape subsequent interactions, and where fairness norms are known to play a key role~\cite{Slembeck1999}. On the biological side, the model speaks to repeated resource-sharing interactions among social animals. Notably, modified UG experiments with chimpanzees have shown that equal reward divisions emerge when obtaining resources requires a partner's cooperation, demonstrating that UG-like allocation problems can capture relevant aspects of non-human social behaviour~\cite{Proctor2013}. Although such experiments have not yet incorporated the reactive memory mechanism studied here, they confirm that repeated dyadic allocation interactions of this type are experimentally realizable in non-human systems, suggesting a natural avenue for future empirical work.

Our model has several limitations that also open avenues for future research. Asynchronous interactions are common in nature, where helping others and receiving help may occur at different time steps. Moreover, in the standard UG, the accepter’s action naturally follows the offerer’s move. In contrast, our model assumes that the offerer and the accepter act simultaneously in each round. In addition, organisms with higher cognitive abilities may engage in repeated interactions using more sophisticated Markovian strategies. Incorporating alternating interactions and complex Markovian strategies into our framework~\cite{NS_1994_JTB, PNH_2022_NATC, PD_2012_PNAS, LHN_2023_PLoS} may therefore capture more realistic social interactions.

An important extension of the present framework concerns population structure. Here, we deliberately considered a well-mixed population in order to isolate the evolutionary consequences of repeated interactions from those of spatial or network reciprocity. In structured populations, local interaction patterns naturally increase the likelihood of repeated encounters and can themselves promote nontrivial evolutionary outcomes in the ultimatum game \cite{SPS_2012_EPL, PNS_2000_PRSB, CWL_2015_EPL, DBA_2016_EHB}. Population structure may therefore interact with the repeated-game mechanism considered here and either reinforce or modify the emergence of fairness. Determining the robustness of the present results on lattices and complex networks constitutes an important direction for future investigation.

Furthermore, when individuals repeatedly interact with their partners, their behaviour is likely to evolve through observation and learning. Experimental studies confirm that learning can help to overcome unfair behaviour~\cite{Roth1995}. A systematic investigation of different kinds of learning rules within our framework remains an important direction for future research~\cite{IT_1965_EBV, CH_1999_E, MF_2002_PNAS, PSPC_2024_NJP}. Moreover, it would be worthwhile to examine scenarios in which individuals occasionally make mistakes in choosing actions or in observing others’ behaviour, particularly in noisy environments~\cite{S_1975_IJGT, AT_2011_EL, IYH_2017_JTB, CAC_2024_PRE}. We conclude with the hope that our work will serve as a catalyst for more experiments to research on repeated interaction driven emergence and sustenance of fairness.

\section*{Data availability statement}
All the relevant numerical codes used to generate the results of this paper are publicly available on GitHub:~\href{https://github.com/ArunavaHub/RepeatedFairness.git}{\url{https://github.com/ArunavaHub/RepeatedFairness.git}}.

\acknowledgements
University Grant Commission (India) is acknowledged for awarding a Senior Research Fellowship to PM.  AP acknowledges the support from the Indian Institute of Technology Kanpur in the form of FARE Fellowship.

\appendix

\section{Analytical derivation of the payoff and fairness rate}\label{appendix:payoff_fairness_rate}
The expected payoffs in Eq.~(\ref{eq:Average_cum_payoff_offerer}) and Eq.~(\ref{eq:Average_cum_payoff_accepter}) can be derived analytically by modelling the repeated interaction as a Markov chain over a finite set of states defined by action pairs---notationally, $\omega=(s_o, s_a)$---of the offerer and the accepter, viz., $(\text{H}, \text{H})$, $(\text{H}, \text{L})$, $(\text{L}, \text{H})$, and $(\text{L}, \text{L})$. Let $\omega=(s_o, s_a)$ and $\omega'=(s'_o, s'_a)$ denote two consecutive outcomes. Then, the transition probability from $\omega$ to $\omega'$ is given by the following elements of transition matrix [say, ${\sf M}(S_o, S_a)$] of the Markov chain:
\begin{equation}
	m_{\omega, \omega'}= z_o\cdot z_a
\end{equation}
where probabilities $z_o$ and $z_a$, respectively, are 
\begin{subequations}
\begin{eqnarray}
	z_o =
	\begin{cases}
		p_o~~~~~~~~~\text{if}~s'_o=\text{H}~\text{and}~s_a=\text{H}, & \\
		1-p_o~~~~ \text{if}~s'_o=\text{L}~\text{and}~s_a=\text{H}, & \\
		q_o~~~~~~~~~\text{if}~s'_o=\text{H}~\text{and}~s_a=\text{L},&\\
		1-q_o~~~~\text{if}~s'_o=\text{L}~\text{and}~s_a=\text{L};
	\end{cases}  \\
	z_a =
	\begin{cases}
		p_a~~~~~~~~~\text{if}~s'_a=\text{H}~\text{and}~s_o=\text{H}, & \\
		1-p_a~~~~ \text{if}~s'_a=\text{L}~\text{and}~s_o=\text{H}, & \\
		q_a~~~~~~~~~\text{if}~s'_a=\text{H}~\text{and}~s_o=\text{L},&\\
		1-q_a~~~~ \text{if}~s'_a=\text{L}~\text{and}~s_o=\text{L}.
	\end{cases}   
\end{eqnarray}
\end{subequations}

Accordingly, ordering the states as $\{\text{HH},\text{HL},\text{LH},\text{LL}\}$, the transition matrix can be written explicitly as follows:
\begin{equation}
	\sf M=
	\begin{pmatrix}
		p_o p_a
		& p_o(1-p_a)
		& (1-p_o)p_a
		& (1-p_o)(1-p_a)
		\\[2mm]
		q_o p_a
		& q_o(1-p_a)
		& (1-q_o)p_a
		& (1-q_o)(1-p_a)
		\\[2mm]
		p_o q_a
		& p_o(1-q_a)
		& (1-p_o)q_a
		& (1-p_o)(1-q_a)
		\\[2mm]
		q_o q_a
		& q_o(1-q_a)
		& (1-q_o)q_a
		& (1-q_o)(1-q_a)
	\end{pmatrix}.
	\label{eq:transition_matrix}
\end{equation}
Here, the rows correspond to the current states $\omega$ and the columns to the subsequent states $\omega'$.

Let ${\bm \sigma}^{(0)}$ denote the initial state vector, where each element $\sigma^{(0)}_{s_o s_a}$ represents the probability that the initial actions of the offerer and the accepter are $s_o$ and $s_a$, respectively. The weighted average state vector is then given by~\cite{HSC_2018_NAT}
\begin{equation}
	{\bm {\bar{\sigma}}}=\frac{\sum_{n=1}^{\infty} {\bm \sigma}^{(0)}(\delta{\sf M})^{n-1}}{\sum_{n=1}^{\infty}\delta^{n-1}}=(1-\delta){\bm \sigma}^{(0)}({\sf I}-\delta{\sf M})^{-1},
	\label{eq:wighted_averaged_state_vector}
\end{equation}
where $\sf I$ is the $4\times4$ identity matrix. The element $\bar{\sigma}_{s_o s_a}$ can be interpreted as the probability of observing the state $\omega=(s_o, s_a)$ over the effective game length. Given the payoff associated with each state, the expected payoff can be obtained by weighting each payoff with $\bar{\sigma}_{s_o s_a}$. Accordingly, the expected payoffs for the offerer and the accepter are, respectively, given by
\begin{subequations}
\begin{eqnarray}
	\pi_o(S_o, S_a)&=&(1-h)\left(\bar{\sigma}_{\text{HH}}+\bar{\sigma}_{\text{HL}}\right)+(1-l)\bar{\sigma}_{\text{LL}},\qquad\\
	\pi_a(S_o, S_a)&=&h\left(\bar{\sigma}_{\text{HH}}+\bar{\sigma}_{\text{HL}}\right)+l\bar{\sigma}_{\text{LL}}.
	\label{eq:wighted_sum_payoff}
\end{eqnarray}
\end{subequations}
Furthermore, the fairness rate corresponding to the strategy pair $(S_o, S_a)$ can be computed from the weighted average state vector as
\begin{equation}
	\gamma(S_o, S_a)=\bar{\sigma}_{\text{HH}}.
	\label{eq:fairness_rate}
\end{equation}%
The value of $\bar{\sigma}_{s_o s_a}$ is, of course, dependent on the strategy pair $(S_o, S_a)$, and the appropriate notation would be $\bar{\sigma}_{s_o s_a}(S_o, S_a)$. However, we retain $\bar{\sigma}_{s_o s_a}$ for notational convenience.

\section{Long-term evolutionary dynamics in two-species population}\label{appendix:long_term_evolution_two_species}
The long-term evolutionary dynamics in a well-mixed, homogeneous population are typically described by the Imhof--Fudenberg--Nowak process~\cite{FI_2006_JET, IN_2009_PRSB}, and it provides a tractable analytical framework for modelling mutation–selection dynamics under rare mutations. This process is widely used to characterize long-run behaviour in single-population settings. We here extend the Imhof--Fudenberg--Nowak process in two-population settings and present an analytical description of the long-term mutation--selection process in a two-species population consisting of offerers and accepters.

In doing so, we consider a finite, well-mixed heterogeneous population under a rare-mutation process, in which each mutation either fixes in the population or goes extinct before another mutation occurs. Consequently, in either case, the population remains monomorphic at each time step and undergoes a sequence of transitions among population states over long evolutionary time scales.

The scenario described above can be fully characterized analytically using a discrete-state, discrete-time Markov chain. The states of the Markov chain correspond to the possible population states, each characterized by a pair of strategies adopted by the two subpopulations. There are 16 possible population states, since the strategy sets of the offerer and the accepter---$\mathcal{S}_o=\{\text{F}_o, \text{A}_o, \text{C}_o, \text{U}_o\}$ and $\mathcal{S}_a=\{\text{F}_a, \text{A}_a, \text{C}_a, \text{U}_a\}$, respectively---have four elements each.

In the mutation–selection process of the heterogeneous population, different mutations in a given resident population can lead to the same population state in the next evolutionary time step. For example, the transition from the population state $(\text{C}_o, \text{C}_a)$ to $(\text{F}_o, \text{C}_a)$ can occur due to the occurrence of four types of mutant pairs: $(\text{F}_o, \text{C}_a)$, $(\text{F}_o, \text{A}_a)$, $(\text{F}_o, \text{U}_a)$, and $(\text{F}_o, \text{F}_a)$. In the first case, the mutation arises in only one subpopulation and must reach fixation. In contrast, in the latter three cases, mutations arise in both subpopulations, but only the mutant in the offerer population must invade in order to reach the final state $(\text{F}_o, \text{C}_a)$.

Thus, the transition probabilities between two monomorphic population states are determined by the fixation probabilities of the mutants that can lead to the same future population state from a given resident population state. We numerically compute these probabilities using the birth--death process in a heterogeneous population (see Appendix~\ref{appendix:birth-death_process}). It is important to note that mutants must appear in both subpopulations in order to generate transitions between population states such as from $(S_o, S_a)$ to $(\tilde{S}_o, \tilde{S}_a)$, with $S_o\neq\tilde{S}_o$ and $S_a\neq\tilde{S}_a$. In this case, the transition probability is determined by the fixation probability of the mutant pair $(\tilde{S}_o, \tilde{S}_a)$.

To express the transition probabilities mathematically, we denote the fixation probability by $\rho_{{S}'_o{S}'_a}(({S}_{o}, {S}_{a}), (\tilde{S}_{o}, \tilde{S}_{a}))$ for a mutant $(\tilde{S}_{o}, \tilde{S}_{a})$ arising in the resident population state $({S}_{o}, {S}_{a})$, which results in the future population state $({S}'_o, {S}'_a)$ where ${S}'_o\in\{S_o, \tilde{S}_o\}$ and ${S}'_a\in\{S_a, \tilde{S}_a\}$. As an illustration, $\rho_{\tilde{S}_{o} {S}_{a}}(({S}_{o}, {S}_{a}), (\tilde{S}_{o}, \tilde{S}_{a}))$ denotes the probability that the mutant pair $(\tilde{S}_{o}, \tilde{S}_{a})$ arising in the resident state $({S}_{o}, {S}_{a})$ leads to transition from $({S}_{o}, {S}_{a})$  to $(\tilde{S}_{o}, {S}_{a})$. In addition, we use $\mu_o$ and $\mu_a$ to represent the probability of mutation occurring in the offerer and the accepter subpopulations, respectively, at each time step.

With all the required elements in place, we now proceed to formulate the transition probabilities in order to construct the Markov chain transition matrix. There are three different types of transitions between the states: $({S}_{o}, {S}_{a})$ to $(\tilde{S}_{o}, {S}_{a})$, $({S}_{o}, {S}_{a})$ to $({S}_{o}, \tilde{S}_{a})$ and $({S}_{o}, {S}_{a})$ to $(\tilde{S}_{o}, \tilde{S}_{a})$. Let us discuss them one by one:
\begin{itemize}
\item The transition probability from the state $({S}_{o}, {S}_{a})$ to the state $ (\tilde{S}_{o}, {S}_{a})$ can be written as
\begin{eqnarray*}
	r[({S}_{o}, {S}_{a}), (\tilde{S}_{o}, {S}_{a})]&=& \mu_o(1-\mu_a)\rho_{\tilde{S}_{o}{S}_{a}}[({S}_{o}, {S}_{a}), (\tilde{S}_{o}, {S}_{a})]\\
	&+& \mu_o\mu_a\sum_{\substack{S'_{a}\neq S_a\\ S_a\in\mathcal{S}_a}}\rho_{\tilde{S}_{o}{S}_{a}}[({S}_{o}, {S}_{a}), (\tilde{S}_{o}, {S}'_{a})].
\end{eqnarray*}
The first term represents the mutation process in which a mutation appears in the offerer subpopulation but does not arise in the accepter subpopulation; obviously, mutation compatible with this process occurs with probability $\mu_o(1-\mu_a)$. The second term represents the mutation process in which mutations appear in both subpopulations but fixation occurs only in the offerer subpopulation. The factor $\mu_o\mu_a$ represents the probability that mutations arise in both subpopulations. Note that $\mu_o\mu_a < \mu_o(1-\mu_a)$ since $\mu_o, \mu_a \ll 1$. This reflects the fact that the simultaneous occurrence of mutations is rarer than the occurrence of a mutation in only one subpopulation.
\item In a similar manner, the expression of the transition probability for the transition $({S}_{o}, {S}_{a})$ to $ ({S}_{o}, \tilde{S}_{a})$ can be written as
\begin{eqnarray*}
	r[({S}_{o}, {S}_{a}), ({S}_{o}, \tilde{S}_{a})]&=& (1-\mu_o)\mu_a\rho_{{S}_{o}\tilde{S}_{a}}[({S}_{o}, {S}_{a}), ({S}_{o}, \tilde{S}_{a})]\\
	&+& \mu_o\mu_a\sum_{\substack{S'_{o}\neq S_o\\ S_o\in\mathcal{S}_o}}\rho_{{S}_{o}\tilde{S}_{a}}[({S}_{o}, {S}_{a}), ({S}'_{o}, \tilde{S}_{a})].
\end{eqnarray*}
\item The transition from state $({S}_{o}, {S}_{a})$ to $(\tilde{S}_{o}, \tilde{S}_{a})$  occurs only if mutations appear and fixate in both subpopulations. Therefore, the transition probability for this case is
\begin{eqnarray*}
	r[({S}_{o}, {S}_{a}), (\tilde{S}_{o}, \tilde{S}_{a})]&=& \mu_o\mu_a\rho_{\tilde{S}_{o}\tilde{S}_{a}}[({S}_{o}, {S}_{a}), (\tilde{S}_{o}, \tilde{S}_{a})].
\end{eqnarray*}
\item Finally, the probability that no transition occurs between the population states can be expressed as
\begin{eqnarray*}
	r[({S}_{o}, {S}_{a}), ({S}_{o}, {S}_{a})]&=& 1-r[({S}_{o}, {S}_{a}), (\tilde{S}_{o}, {S}_{a})]- r[({S}_{o}, {S}_{a}), \\
	&&({S}_{o}, \tilde{S}_{a})]-r[({S}_{o}, {S}_{a}), (\tilde{S}_{o}, \tilde{S}_{a})].
\end{eqnarray*}
\end{itemize}

These transition probabilities form the 16 dimensional Markov transition matrix, which we denote by $\sf R$. Let $\bm{\alpha}(t)$ denote the probability row vector at time $t$, whose component $\alpha_{({S}_{o}, {S}_{a})}^{(t)}$  represents the abundance of the population state $(S_o, S_a)$ at time $t$. This probability vector is obtained using the Markov chain equation $\bm{\alpha}(t)=\bm{\alpha}(0){\sf R}^t$, given an initial probability vector $\bm{\alpha}(0)$. As we consider the initial state to be the unfair strategy pair $(\text{U}_o, \text{U}_a)$, we assign 1 to the component of $\bm{\alpha}(0)$ corresponding to the state $(\text{U}_o, \text{U}_a)$ and 0 to other components.

\section{Birth-death process in two-species population}
\label{appendix:birth-death_process}
In this section, we present the birth--death process in a heterogeneous population in order to determine the fixation probability—the probability with which a mutant replaces a resident population—required to formulate the Markov chain framework for stochastic long-term evolutionary dynamics. We, therefore, consider a well-mixed heterogeneous population consisting of two subpopulations: offerers and accepters, whose population sizes are denoted by $N_o$ and $N_a$, respectively. We assume that each offerer can adopt one of two possible strategies, $S_o$ and $\tilde{S}_o$, and each accepter can adopt one of the two available strategies, $S_a$ and $\tilde{S}_a$. Furthermore, we assume that the strategic interaction between the offerer and the accepter subpopulations is governed by the bimatrix game presented in Eq.~(\ref{eq:Repeated_game_matrix}). The state of the heterogeneous population at any point in time is denoted by $(n_o, n_a)$, where $n_o$ is the number of offerers with strategy $\tilde{S}_o$ in the offerer subpopulation and $n_a$ is the number of accepters with strategy $\tilde{S}_a$ in the accepter subpopulation. 

A offerer accumulates an expected payoff by interacting with $(N_a-n_a)$ accepters with strategy $S_a$ and $n_a$ accepters with strategy $\tilde{S}_a$, whereas an accepter receives an expected payoff from interactions with $(N_o-n_o)$ offerers with strategy $S_o$ and $n_o$ offerers with strategy $\tilde{S}_o$. Therefore, the expected payoffs of an offerer with strategies $S_o$ and $\tilde{S}_o$ can be written as
\begin{eqnarray*}
	E_{S_o}(n_a)&=&\frac{1}{N_a}\left[(N_a-n_a)\pi_o(S_o, S_a)+ n_a\pi_o(S_o, \tilde{S}_a)\right],\\
	E_{\tilde{S}_o}(n_a)&=&\frac{1}{N_a}\left[(N_a-n_a)\pi_o(\tilde{S}_o, S_a)+n_a\pi_o(\tilde{S}_o, \tilde{S}_a)\right].
\end{eqnarray*}
Likewise, the expected payoffs of an accepter with strategies $S_a$ and $\tilde{S}_a$ are given by
\begin{eqnarray*}
	E_{S_a}(n_o)&=&\frac{1}{N_o}\left[(N_o-n_o)\pi_a(S_o, S_a)+n_o\pi_a(\tilde{S}_o, S_a)\right],\\
	E_{\tilde{S}_a}(n_o)&=&\frac{1}{N_o}\left[(N_o-n_o)\pi_a(S_o, \tilde{S}_a)+n_o\pi_a(\tilde{S}_o, \tilde{S}_a)\right].
\end{eqnarray*}%

We now define the fitness of a player as an exponential function~\cite{RTON_2013_PNAS, SO_2015_DGA} of the expected payoff of that player. Therefore, the fitness of $S_o$ and $\tilde{S}_o$ strategist offerers are $$f_{S_o}=e^{wE_{S_o}}~~\text{and}~~ f_{\tilde{S}_o}=e^{wE_{\tilde{S}_o}};$$ in a similar manner, the fitness of $S_a$ and $\tilde{S}_a$ strategist accepters are given by, $$f_{S_a}=e^{wE_{S_a}}~~\text{and}~~ f_{\tilde{S}_a}=e^{wE_{\tilde{S}_a}}.$$  
Here, $w$ represents the intensity of selection, which can take values from zero to infinity. $w\rightarrow 0$ represents the dominance of neutral drift. As $w$ increases, the role of payoffs in fitness increases, and therefore selection becomes stronger. 

We employ the birth--death process to govern the replication--selection dynamics of the strategies in the population, and numerically determine the fixation probabilities. At each time step, one individual from each subpopulation is selected for reproduction with probability proportional to her fitness. Then, one individual from each subpopulation is randomly chosen to die. The offspring in each subpopulation subsequently replaces the individual that dies in the corresponding subpopulation. As a result, the population size in each subpopulation remains constant.

The relative fitness of an offerer with the strategy $S_o$ and $\tilde{S}_o$ are respectively given by,
$$
p^r_{S_o}=\frac{(N_o-n_o)f_{S_o}}{\bar{f}_o}~~\text{and}~~
p^r_{\tilde{S}_o}=\frac{n_of_{\tilde{S}_o}}{\bar{f}_o},
$$
where $\bar{f}_o=(N_o-n_o)f_{S_o}+ n_of_{\tilde{S}_o}$ represents {\color{black}the total fitness} of the offerer subpopulation.
Similarly, the relative fitness of an accepter with strategy $S_a$ and $\tilde{S}_a$ can be respectively calculated as 
$$
p^r_{S_a}=\frac{(N_a-n_a)f_{S_a}}{\bar{f}_a}~~\text{and}~~
p^r_{\tilde{S}_a}=\frac{n_af_{\tilde{S}_a}}{\bar{f}_a},
$$
where $\bar{f}_a=(N_a-n_a)f_{S_a}+ n_af_{\tilde{S}_a}$ represents  {\color{black}the total fitness} of the accepter subpopulation. The superscript $r$ is used to represent the probability of a player being chosen for reproduction. 

Subsequently, we determine the probability that a player of a given type is chosen for death, which is equal to the frequency of that type in the population. Given the state of the population $(n_o, n_a)$, the probability that an offerer with strategy $\tilde{S}_o$ and an accepter with strategy $\tilde{S}_a$ are selected for death are given by
$$
 p^d_{\tilde{S}_o}=\frac{n_o}{N_o}~~\text{and}~~
 p^d_{\tilde{S}_a}=\frac{n_a}{N_a}.
$$ 
Clearly, the probabilities that an offerer with strategy $S_o$ and an accepter with strategy $S_a$ are selected for death are $p^d_{S_o}=1-p^d_{\tilde{S}o}$ and $p^d_{S_a}=1-p^d_{\tilde{S}_a}$. Here, the superscript $d$ denotes the probability that a player is selected to die. These probabilities govern the stochastic short-term evolutionary dynamics of the strategies in the heterogeneous population.

To numerically determine the fixation probabilities, we introduce one mutant in either or both subpopulations, depending on which fixation probability needs to be computed, and run the process for a sufficiently long time until the mutant either fixates or goes extinct. In the case where mutants arise in both subpopulations, we determine the fixation probabilities in the following way. Suppose mutants with strategy pair $(\tilde{S}_o, \tilde{S}_a)$ arise in a heterogeneous population with resident strategy pair $(S_o, S_a)$. As discussed earlier, this situation can generate three distinct new population states: $(N_o, 0)$, $(0, N_a)$, and $(N_o, N_a)$. Therefore, we examine which of these three states the population is absorbed into, starting from the initial distribution $(n_o,n_a)=(1,1)$. Since the process is stochastic in nature, we repeat the process for multiple ($10^5$) trials starting from the same initial distribution $(n_o,n_a)=(1,1)$. We then record the frequency of trials---each run up to $10^4$ time steps---in which the population is absorbed in $(N_o, 0)$, $(0, N_a)$, and $(N_o, N_a)$ in order to determine the fixation probabilities $\rho_{\tilde{S}_{o}S_{a}}[({S}_{o}, {S}_{a}), (\tilde{S}_{o}, \tilde{S}_{a})]$, $\rho_{S_{o}\tilde{S}_{a}}[({S}_{o}, {S}_{a}), (\tilde{S}_{o}, \tilde{S}_{a})]$ and $\rho_{\tilde{S}_{o}\tilde{S}_{a}}[({S}_{o}, {S}_{a}), (\tilde{S}_{o}, \tilde{S}_{a})]$, respectively. In passing, we mention that an analytical expression for fixation probabilities in a two-species population exists in the literature, but it is valid only in the weak selection limit~\cite{SO_2015_DGA}.
 
The remaining two fixation probabilities $\rho_{\tilde{S}_{o}S_{a}}[({S}_{o}, {S}_{a}), (\tilde{S}_{o}, S_{a})]$ and $\rho_{S_{o} \tilde{S}_{a}}[({S}_{o}, {S}_{a}), (S_{o}, \tilde{S}_{a})]$ can be obtained by introducing the mutant in one subpopulation. In such a case, the mutant gives rise to an unique final state upon fixation. For example, a mutant $(\tilde{S}_{o},S_{a})$ in the resident population state $(S_o, S_a)$ can reach the final state $(\tilde{S}_{o},S_{a})$ upon fixation. Therefore, the fixation probability is the frequency of trials that are absorbed in the state $(N_o,0)$ starting from $(1,0)$. This situation is similar to the birth--death process in a homogeneous population~\cite{N_2006_BOOK}, since the state of the subpopulation in which no mutation occurs remains unchanged over time. It allows us to use the analytical expression for the fixation probability in homogeneous population with appropriate modifications to find the fixation probability $\rho_{\tilde{S}_{o}S_{a}}[({S}_{o}, {S}_{a}), (\tilde{S}_{o}, S_{a})]$. Similarly, we find the fixation probability $\rho_{S_{o} \tilde{S}_{a}}[({S}_{o}, {S}_{a}), (S_{o}, \tilde{S}_{a})]$.

\begin{figure}[h!]
		\centering
		\includegraphics[scale=0.4]{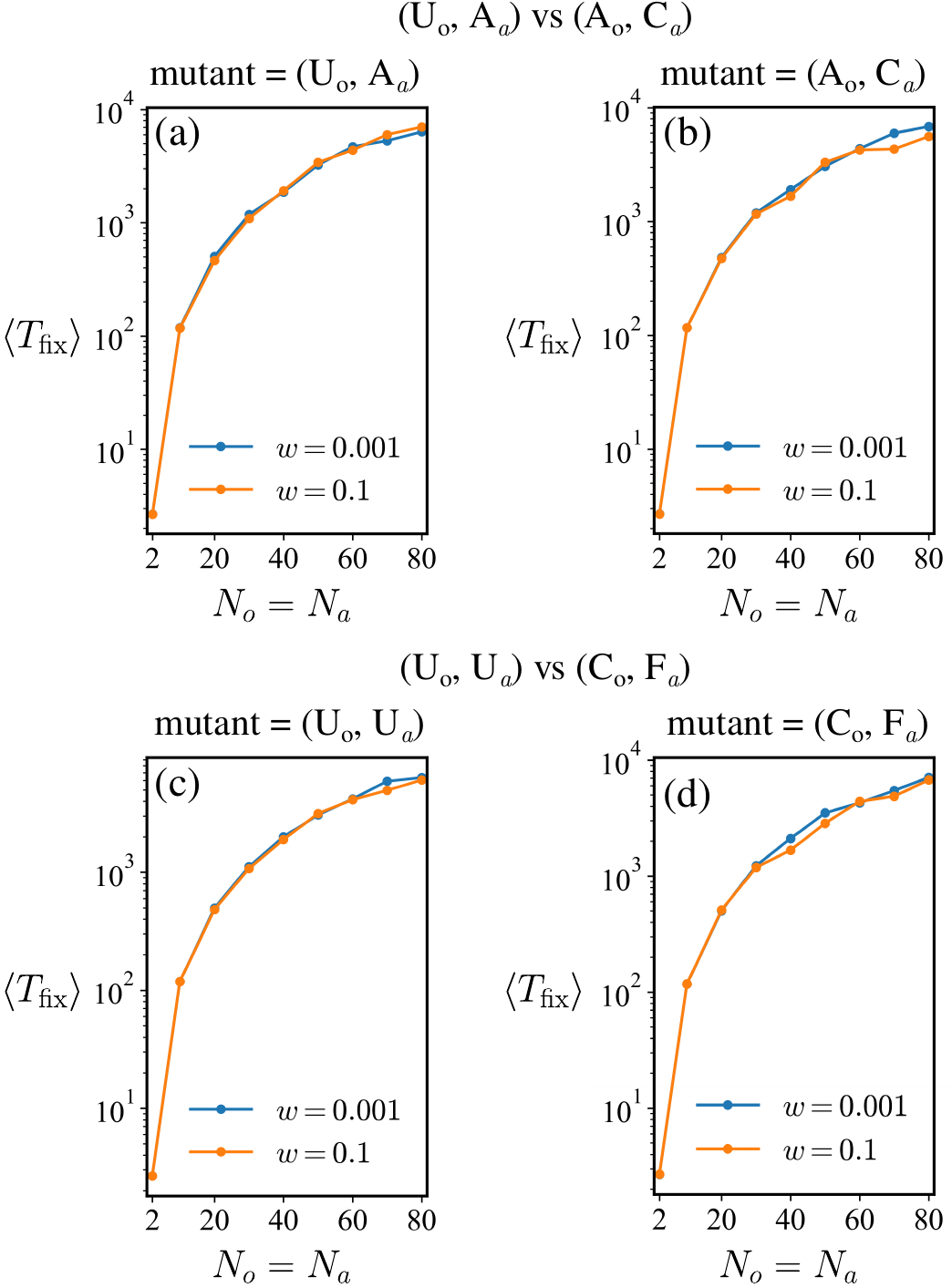}
		\caption{Fixation time as a function of population size.  The blue and the orange curves depict the fixation times for selection strengths $w=0.001$ and $w=0.1$, respectively.}	
		\label{fig:fixation_popu}
	\end{figure}

In the above calculation of the fixation probability, we find that running the numerical code for $10^4$ time steps suffices to find the fixation probability	for the population-sizes we have worked with. In fact, this conclusion is independent of the kind of selection strength we have chosen; there is even no practical enhancement of fixation time when compared with low selection strength. To illustrate these facts, we consider two combinations of resident--mutant:
\begin{enumerate}
	\item $(\text{U}_o,\text{A}_a)$ and $(\text{A}_o,\text{C}_a)$ such that in terms of payoffs with $\delta=0.9$, they appear to be mutually invading each other, i.e., $$\pi_o(\text{U}_o,\text{C}_a)> \pi_o(\text{A}_o,\text{C}_a),~~~\pi_a(\text{A}_o,\text{A}_a)>\pi_a(\text{A}_o,\text{C}_a);$$
		$$\pi_o(\text{A}_o,\text{A}_a)>\pi_o(\text{U}_o,\text{A}_a),~~~\pi_a(\text{U}_o,\text{C}_a)>\pi_a(\text{U}_o,\text{A}_a).$$
	\item $(\text{U}_o,\text{U}_a)$ and $(\text{C}_o,\text{F}_a)$ such that in terms of payoffs  with $\delta=0.9$, neither appear to be invading the other, i.e., $$\pi_o(\text{U}_o,\text{F}_a)<\pi_o(\text{C}_o,\text{F}_a),~~~\pi_a(\text{C}_o,\text{U}_a)<\pi_a(\text{C}_o,\text{F}_a);$$ $$\pi_o(\text{C}_o,\text{U}_a)<\pi_o(\text{U}_o,\text{U}_a),~~~\pi_a(\text{U}_o,\text{F}_a)<\pi_a(\text{U}_o,\text{U}_a).$$
\end{enumerate}
For the two cases above, the mean fixation time, $\langle T_{\text{fix}}\rangle$, is estimated and presented in Fig.~\ref{fig:fixation_popu} for different population sizes. In both the cases, as expected, $\langle T_{\text{fix}}\rangle$ increases with the population size, although the rate of increase becomes slower for larger populations. The fixation time increases by approximately a factor of $10^4$ as the population size increases from $N_o=N_a=2$ to $N_o=N_a=80$. Furthermore, the fixation times are nearly identical for the low selection strength ($w=0.001$) and the higher selection strength ($w=0.1$), as have been depicted, respectively, by the blue and the orange curves in Fig.~\ref{fig:fixation_popu}. We again note here that it suffices to run the simulations for $10^4$ time steps to observe fixations. 
\bibliography{Mandal_etal_manuscript_reference}

\begin{thebibliography}{79}%
\makeatletter
\providecommand \@ifxundefined [1]{%
 \@ifx{#1\undefined}
}%
\providecommand \@ifnum [1]{%
 \ifnum #1\expandafter \@firstoftwo
 \else \expandafter \@secondoftwo
 \fi
}%
\providecommand \@ifx [1]{%
 \ifx #1\expandafter \@firstoftwo
 \else \expandafter \@secondoftwo
 \fi
}%
\providecommand \natexlab [1]{#1}%
\providecommand \enquote  [1]{``#1''}%
\providecommand \bibnamefont  [1]{#1}%
\providecommand \bibfnamefont [1]{#1}%
\providecommand \citenamefont [1]{#1}%
\providecommand \href@noop [0]{\@secondoftwo}%
\providecommand \href [0]{\begingroup \@sanitize@url \@href}%
\providecommand \@href[1]{\@@startlink{#1}\@@href}%
\providecommand \@@href[1]{\endgroup#1\@@endlink}%
\providecommand \@sanitize@url [0]{\catcode `\\12\catcode `\$12\catcode
  `\&12\catcode `\#12\catcode `\^12\catcode `\_12\catcode `\%12\relax}%
\providecommand \@@startlink[1]{}%
\providecommand \@@endlink[0]{}%
\providecommand \url  [0]{\begingroup\@sanitize@url \@url }%
\providecommand \@url [1]{\endgroup\@href {#1}{\urlprefix }}%
\providecommand \urlprefix  [0]{URL }%
\providecommand \Eprint [0]{\href }%
\providecommand \doibase [0]{https://doi.org/}%
\providecommand \selectlanguage [0]{\@gobble}%
\providecommand \bibinfo  [0]{\@secondoftwo}%
\providecommand \bibfield  [0]{\@secondoftwo}%
\providecommand \translation [1]{[#1]}%
\providecommand \BibitemOpen [0]{}%
\providecommand \bibitemStop [0]{}%
\providecommand \bibitemNoStop [0]{.\EOS\space}%
\providecommand \EOS [0]{\spacefactor3000\relax}%
\providecommand \BibitemShut  [1]{\csname bibitem#1\endcsname}%
\let\auto@bib@innerbib\@empty
\bibitem [{\citenamefont {Milinski}(2010)}]{M_2010_ZT}%
  \BibitemOpen
  \bibfield  {author} {\bibinfo {author} {\bibfnamefont {M.}~\bibnamefont
  {Milinski}},\ }\bibfield  {title} {\bibinfo {title} {An evolutionarily stable
  feeding strategy in sticklebacks1},\ }\href
  {https://doi.org/10.1111/j.1439-0310.1979.tb00669.x} {\bibfield  {journal}
  {\bibinfo  {journal} {Zeitschrift f\"{u}r Tierpsychologie}\ }\textbf
  {\bibinfo {volume} {51}},\ \bibinfo {pages} {36–40} (\bibinfo {year}
  {2010})}\BibitemShut {NoStop}%
\bibitem [{\citenamefont {Brosnan}\ and\ \citenamefont
  {de~Waal}(2003)}]{BW_2003_NAT}%
  \BibitemOpen
  \bibfield  {author} {\bibinfo {author} {\bibfnamefont {S.~F.}\ \bibnamefont
  {Brosnan}}\ and\ \bibinfo {author} {\bibfnamefont {F.~B.~M.}\ \bibnamefont
  {de~Waal}},\ }\bibfield  {title} {\bibinfo {title} {Monkeys reject unequal
  pay},\ }\href {https://doi.org/10.1038/nature01963} {\bibfield  {journal}
  {\bibinfo  {journal} {Nature}\ }\textbf {\bibinfo {volume} {425}},\ \bibinfo
  {pages} {297–299} (\bibinfo {year} {2003})}\BibitemShut {NoStop}%
\bibitem [{\citenamefont {Brosnan}\ \emph {et~al.}(2005)\citenamefont
  {Brosnan}, \citenamefont {Schiff},\ and\ \citenamefont
  {de~Waal}}]{BSW_2005_PRSB}%
  \BibitemOpen
  \bibfield  {author} {\bibinfo {author} {\bibfnamefont {S.~F.}\ \bibnamefont
  {Brosnan}}, \bibinfo {author} {\bibfnamefont {H.~C.}\ \bibnamefont
  {Schiff}},\ and\ \bibinfo {author} {\bibfnamefont {F.~B.~M.}\ \bibnamefont
  {de~Waal}},\ }\bibfield  {title} {\bibinfo {title} {Tolerance for inequity
  may increase with social closeness in chimpanzees},\ }\href
  {https://doi.org/10.1098/rspb.2004.2947} {\bibfield  {journal} {\bibinfo
  {journal} {Proceedings of the Royal Society B: Biological Sciences}\ }\textbf
  {\bibinfo {volume} {272}},\ \bibinfo {pages} {253–258} (\bibinfo {year}
  {2005})}\BibitemShut {NoStop}%
\bibitem [{\citenamefont {Range}\ \emph {et~al.}(2009)\citenamefont {Range},
  \citenamefont {Horn}, \citenamefont {Viranyi},\ and\ \citenamefont
  {Huber}}]{RHVH_2009_PNAS}%
  \BibitemOpen
  \bibfield  {author} {\bibinfo {author} {\bibfnamefont {F.}~\bibnamefont
  {Range}}, \bibinfo {author} {\bibfnamefont {L.}~\bibnamefont {Horn}},
  \bibinfo {author} {\bibfnamefont {Z.}~\bibnamefont {Viranyi}},\ and\ \bibinfo
  {author} {\bibfnamefont {L.}~\bibnamefont {Huber}},\ }\bibfield  {title}
  {\bibinfo {title} {The absence of reward induces inequity aversion in dogs},\
  }\href {https://doi.org/10.1073/pnas.0810957105} {\bibfield  {journal}
  {\bibinfo  {journal} {Proceedings of the National Academy of Sciences}\
  }\textbf {\bibinfo {volume} {106}},\ \bibinfo {pages} {340–345} (\bibinfo
  {year} {2009})}\BibitemShut {NoStop}%
\bibitem [{\citenamefont {Bekoff}(2004)}]{B_2004_BP}%
  \BibitemOpen
  \bibfield  {author} {\bibinfo {author} {\bibfnamefont {M.}~\bibnamefont
  {Bekoff}},\ }\bibfield  {title} {\bibinfo {title} {Wild justice and fair
  play: cooperation, forgiveness, and morality in animals},\ }\href
  {https://doi.org/10.1007/sbiph-004-0539-x} {\bibfield  {journal} {\bibinfo
  {journal} {Biology \& amp; Philosophy}\ }\textbf {\bibinfo {volume} {19}},\
  \bibinfo {pages} {489–520} (\bibinfo {year} {2004})}\BibitemShut {NoStop}%
\bibitem [{\citenamefont {Heinrich}(2009)}]{H_1999_BOOK}%
  \BibitemOpen
  \bibfield  {author} {\bibinfo {author} {\bibfnamefont {B.}~\bibnamefont
  {Heinrich}},\ }\href {https://books.google.co.in/books?id=D2IlvE7uDZkC}
  {\emph {\bibinfo {title} {Mind of the Raven: Investigations and Adventures
  with Wolf-Birds}}}\ (\bibinfo  {publisher} {HarperCollins},\ \bibinfo {year}
  {2009})\BibitemShut {NoStop}%
\bibitem [{\citenamefont {Brosnan}(2006)}]{B_2006_SJR}%
  \BibitemOpen
  \bibfield  {author} {\bibinfo {author} {\bibfnamefont {S.~F.}\ \bibnamefont
  {Brosnan}},\ }\bibfield  {title} {\bibinfo {title} {Nonhuman species’
  reactions to inequity and their implications for fairness},\ }\href@noop {}
  {\bibfield  {journal} {\bibinfo  {journal} {Social Justice Research}\
  }\textbf {\bibinfo {volume} {19}},\ \bibinfo {pages} {153–185} (\bibinfo
  {year} {2006})}\BibitemShut {NoStop}%
\bibitem [{\citenamefont {Fehr}\ and\ \citenamefont
  {Schmidt}(1999)}]{FS_1999_QJE}%
  \BibitemOpen
  \bibfield  {author} {\bibinfo {author} {\bibfnamefont {E.}~\bibnamefont
  {Fehr}}\ and\ \bibinfo {author} {\bibfnamefont {K.~M.}\ \bibnamefont
  {Schmidt}},\ }\bibfield  {title} {\bibinfo {title} {A theory of fairness,
  competition, and cooperation},\ }\href {http://www.jstor.org/stable/2586885}
  {\bibfield  {journal} {\bibinfo  {journal} {The Quarterly Journal of
  Economics}\ }\textbf {\bibinfo {volume} {114}},\ \bibinfo {pages} {817}
  (\bibinfo {year} {1999})}\BibitemShut {NoStop}%
\bibitem [{\citenamefont {Fehr}\ and\ \citenamefont
  {Fischbacher}(2003)}]{FF_2003_NAT}%
  \BibitemOpen
  \bibfield  {author} {\bibinfo {author} {\bibfnamefont {E.}~\bibnamefont
  {Fehr}}\ and\ \bibinfo {author} {\bibfnamefont {U.}~\bibnamefont
  {Fischbacher}},\ }\bibfield  {title} {\bibinfo {title} {The nature of human
  altruism},\ }\href {https://doi.org/10.1038/nature02043} {\bibfield
  {journal} {\bibinfo  {journal} {Nature}\ }\textbf {\bibinfo {volume} {425}},\
  \bibinfo {pages} {785–791} (\bibinfo {year} {2003})}\BibitemShut {NoStop}%
\bibitem [{\citenamefont {André}\ and\ \citenamefont
  {Baumard}(2011)}]{AB_2011_JTB}%
  \BibitemOpen
  \bibfield  {author} {\bibinfo {author} {\bibfnamefont {J.-B.}\ \bibnamefont
  {André}}\ and\ \bibinfo {author} {\bibfnamefont {N.}~\bibnamefont
  {Baumard}},\ }\bibfield  {title} {\bibinfo {title} {Social opportunities and
  the evolution of fairness},\ }\href
  {https://doi.org/10.1016/j.jtbi.2011.07.031} {\bibfield  {journal} {\bibinfo
  {journal} {Journal of Theoretical Biology}\ }\textbf {\bibinfo {volume}
  {289}},\ \bibinfo {pages} {128–135} (\bibinfo {year} {2011})}\BibitemShut
  {NoStop}%
\bibitem [{\citenamefont {Binmore}(2005)}]{B_2005_NJ}%
  \BibitemOpen
  \bibfield  {author} {\bibinfo {author} {\bibfnamefont {K.}~\bibnamefont
  {Binmore}},\ }\href@noop {} {\emph {\bibinfo {title} {Natural Justice}}}\
  (\bibinfo  {publisher} {Oxford University Press},\ \bibinfo {address} {New
  York},\ \bibinfo {year} {2005})\BibitemShut {NoStop}%
\bibitem [{\citenamefont {Brosnan}\ and\ \citenamefont
  {de~Waal}(2014)}]{BW_2014_Sc}%
  \BibitemOpen
  \bibfield  {author} {\bibinfo {author} {\bibfnamefont {S.~F.}\ \bibnamefont
  {Brosnan}}\ and\ \bibinfo {author} {\bibfnamefont {F.~B.~M.}\ \bibnamefont
  {de~Waal}},\ }\bibfield  {title} {\bibinfo {title} {Evolution of responses to
  (un)fairness},\ }\href {http://dx.doi.org/10.1126/science.1251776} {\bibfield
   {journal} {\bibinfo  {journal} {Science}\ }\textbf {\bibinfo {volume} {346}}
  (\bibinfo {year} {2014})}\BibitemShut {NoStop}%
\bibitem [{\citenamefont {Debove}\ \emph {et~al.}(2017)\citenamefont {Debove},
  \citenamefont {Baumard},\ and\ \citenamefont {André}}]{DBA_2017_PO}%
  \BibitemOpen
  \bibfield  {author} {\bibinfo {author} {\bibfnamefont {S.}~\bibnamefont
  {Debove}}, \bibinfo {author} {\bibfnamefont {N.}~\bibnamefont {Baumard}},\
  and\ \bibinfo {author} {\bibfnamefont {J.-B.}\ \bibnamefont {André}},\
  }\bibfield  {title} {\bibinfo {title} {On the evolutionary origins of
  equity},\ }\href {https://doi.org/10.1371/journal.pone.0173636} {\bibfield
  {journal} {\bibinfo  {journal} {PLOS ONE}\ }\textbf {\bibinfo {volume}
  {12}},\ \bibinfo {pages} {e0173636} (\bibinfo {year} {2017})}\BibitemShut
  {NoStop}%
\bibitem [{\citenamefont {Maynard~Smith}\ and\ \citenamefont
  {Price}(1973)}]{SP_1973_NAT}%
  \BibitemOpen
  \bibfield  {author} {\bibinfo {author} {\bibfnamefont {J.}~\bibnamefont
  {Maynard~Smith}}\ and\ \bibinfo {author} {\bibfnamefont {G.~R.}\ \bibnamefont
  {Price}},\ }\bibfield  {title} {\bibinfo {title} {The logic of animal
  conflict},\ }\href {https://doi.org/10.1038/246015a0} {\bibfield  {journal}
  {\bibinfo  {journal} {Nature}\ }\textbf {\bibinfo {volume} {246}},\ \bibinfo
  {pages} {15–18} (\bibinfo {year} {1973})}\BibitemShut {NoStop}%
\bibitem [{\citenamefont {Maynard~Smith}(1974)}]{M_1974_JTB}%
  \BibitemOpen
  \bibfield  {author} {\bibinfo {author} {\bibfnamefont {J.}~\bibnamefont
  {Maynard~Smith}},\ }\bibfield  {title} {\bibinfo {title} {The theory of games
  and the evolution of animal conflicts},\ }\href
  {https://doi.org/10.1016/0022-5193(74)90110-6} {\bibfield  {journal}
  {\bibinfo  {journal} {Journal of Theoretical Biology}\ }\textbf {\bibinfo
  {volume} {47}},\ \bibinfo {pages} {209–221} (\bibinfo {year}
  {1974})}\BibitemShut {NoStop}%
\bibitem [{\citenamefont {Maynard~Smith}(1988)}]{S_1988_BOOK}%
  \BibitemOpen
  \bibfield  {author} {\bibinfo {author} {\bibfnamefont {J.}~\bibnamefont
  {Maynard~Smith}},\ }\href {https://doi.org/10.1007/978-1-4684-7862-4_22}
  {\emph {\bibinfo {title} {Did Darwin Get It Right?}}}\ (\bibinfo  {publisher}
  {Springer US},\ \bibinfo {year} {1988})\ p.\ \bibinfo {pages}
  {202–215}\BibitemShut {NoStop}%
\bibitem [{\citenamefont {Nowak}\ \emph {et~al.}(2000)\citenamefont {Nowak},
  \citenamefont {Page},\ and\ \citenamefont {Sigmund}}]{NPS_2000_Sc}%
  \BibitemOpen
  \bibfield  {author} {\bibinfo {author} {\bibfnamefont {M.~A.}\ \bibnamefont
  {Nowak}}, \bibinfo {author} {\bibfnamefont {K.~M.}\ \bibnamefont {Page}},\
  and\ \bibinfo {author} {\bibfnamefont {K.}~\bibnamefont {Sigmund}},\
  }\bibfield  {title} {\bibinfo {title} {Fairness versus reason in the
  ultimatum game},\ }\href {https://doi.org/10.1126/science.289.5485.1773}
  {\bibfield  {journal} {\bibinfo  {journal} {Science}\ }\textbf {\bibinfo
  {volume} {289}},\ \bibinfo {pages} {1773–1775} (\bibinfo {year}
  {2000})}\BibitemShut {NoStop}%
\bibitem [{\citenamefont {Yang}\ \emph {et~al.}(2015)\citenamefont {Yang},
  \citenamefont {Li}, \citenamefont {Wu},\ and\ \citenamefont
  {Wang}}]{YLW_2015_EL}%
  \BibitemOpen
  \bibfield  {author} {\bibinfo {author} {\bibfnamefont {Z.}~\bibnamefont
  {Yang}}, \bibinfo {author} {\bibfnamefont {Z.}~\bibnamefont {Li}}, \bibinfo
  {author} {\bibfnamefont {T.}~\bibnamefont {Wu}},\ and\ \bibinfo {author}
  {\bibfnamefont {L.}~\bibnamefont {Wang}},\ }\bibfield  {title} {\bibinfo
  {title} {Effects of partner choice and role assignation in the spatial
  ultimatum game},\ }\href {https://doi.org/10.1209/0295-5075/109/40013}
  {\bibfield  {journal} {\bibinfo  {journal} {EPL (Europhysics Letters)}\
  }\textbf {\bibinfo {volume} {109}},\ \bibinfo {pages} {40013} (\bibinfo
  {year} {2015})}\BibitemShut {NoStop}%
\bibitem [{\citenamefont {Henrich}\ \emph {et~al.}(2006)\citenamefont
  {Henrich}, \citenamefont {McElreath}, \citenamefont {Barr}, \citenamefont
  {Ensminger}, \citenamefont {Barrett}, \citenamefont {Bolyanatz},
  \citenamefont {Cardenas}, \citenamefont {Gurven}, \citenamefont {Gwako},
  \citenamefont {Henrich}, \citenamefont {Lesorogol}, \citenamefont {Marlowe},
  \citenamefont {Tracer},\ and\ \citenamefont {Ziker}}]{HMB_2006_Sc}%
  \BibitemOpen
  \bibfield  {author} {\bibinfo {author} {\bibfnamefont {J.}~\bibnamefont
  {Henrich}}, \bibinfo {author} {\bibfnamefont {R.}~\bibnamefont {McElreath}},
  \bibinfo {author} {\bibfnamefont {A.}~\bibnamefont {Barr}}, \bibinfo {author}
  {\bibfnamefont {J.}~\bibnamefont {Ensminger}}, \bibinfo {author}
  {\bibfnamefont {C.}~\bibnamefont {Barrett}}, \bibinfo {author} {\bibfnamefont
  {A.}~\bibnamefont {Bolyanatz}}, \bibinfo {author} {\bibfnamefont {J.~C.}\
  \bibnamefont {Cardenas}}, \bibinfo {author} {\bibfnamefont {M.}~\bibnamefont
  {Gurven}}, \bibinfo {author} {\bibfnamefont {E.}~\bibnamefont {Gwako}},
  \bibinfo {author} {\bibfnamefont {N.}~\bibnamefont {Henrich}}, \bibinfo
  {author} {\bibfnamefont {C.}~\bibnamefont {Lesorogol}}, \bibinfo {author}
  {\bibfnamefont {F.}~\bibnamefont {Marlowe}}, \bibinfo {author} {\bibfnamefont
  {D.}~\bibnamefont {Tracer}},\ and\ \bibinfo {author} {\bibfnamefont
  {J.}~\bibnamefont {Ziker}},\ }\bibfield  {title} {\bibinfo {title} {Costly
  punishment across human societies},\ }\href
  {https://doi.org/10.1126/science.1127333} {\bibfield  {journal} {\bibinfo
  {journal} {Science}\ }\textbf {\bibinfo {volume} {312}},\ \bibinfo {pages}
  {1767–1770} (\bibinfo {year} {2006})}\BibitemShut {NoStop}%
\bibitem [{\citenamefont {Han}\ \emph {et~al.}(2017)\citenamefont {Han},
  \citenamefont {Cao}, \citenamefont {Shen}, \citenamefont {Zhang},
  \citenamefont {Wang}, \citenamefont {Cressman},\ and\ \citenamefont
  {Stanley}}]{HCS_2017_PNAS}%
  \BibitemOpen
  \bibfield  {author} {\bibinfo {author} {\bibfnamefont {X.}~\bibnamefont
  {Han}}, \bibinfo {author} {\bibfnamefont {S.}~\bibnamefont {Cao}}, \bibinfo
  {author} {\bibfnamefont {Z.}~\bibnamefont {Shen}}, \bibinfo {author}
  {\bibfnamefont {B.}~\bibnamefont {Zhang}}, \bibinfo {author} {\bibfnamefont
  {W.-X.}\ \bibnamefont {Wang}}, \bibinfo {author} {\bibfnamefont
  {R.}~\bibnamefont {Cressman}},\ and\ \bibinfo {author} {\bibfnamefont
  {H.~E.}\ \bibnamefont {Stanley}},\ }\bibfield  {title} {\bibinfo {title}
  {Emergence of communities and diversity in social networks},\ }\href
  {https://doi.org/10.1073/pnas.1608164114} {\bibfield  {journal} {\bibinfo
  {journal} {Proceedings of the National Academy of Sciences}\ }\textbf
  {\bibinfo {volume} {114}},\ \bibinfo {pages} {2887–2891} (\bibinfo {year}
  {2017})}\BibitemShut {NoStop}%
\bibitem [{\citenamefont {Debove}\ \emph {et~al.}(2016)\citenamefont {Debove},
  \citenamefont {Baumard},\ and\ \citenamefont {André}}]{DBA_2016_EHB}%
  \BibitemOpen
  \bibfield  {author} {\bibinfo {author} {\bibfnamefont {S.}~\bibnamefont
  {Debove}}, \bibinfo {author} {\bibfnamefont {N.}~\bibnamefont {Baumard}},\
  and\ \bibinfo {author} {\bibfnamefont {J.-B.}\ \bibnamefont {André}},\
  }\bibfield  {title} {\bibinfo {title} {Models of the evolution of fairness in
  the ultimatum game: a review and classification},\ }\href
  {https://doi.org/10.1016/j.evolhumbehav.2016.01.001} {\bibfield  {journal}
  {\bibinfo  {journal} {Evolution and Human Behavior}\ }\textbf {\bibinfo
  {volume} {37}},\ \bibinfo {pages} {245–254} (\bibinfo {year}
  {2016})}\BibitemShut {NoStop}%
\bibitem [{\citenamefont {Henrich}\ \emph {et~al.}(2005)\citenamefont
  {Henrich}, \citenamefont {Boyd}, \citenamefont {Bowles}, \citenamefont
  {Camerer}, \citenamefont {Fehr}, \citenamefont {Gintis}, \citenamefont
  {McElreath}, \citenamefont {Alvard}, \citenamefont {Barr}, \citenamefont
  {Ensminger}, \citenamefont {Henrich}, \citenamefont {Hill}, \citenamefont
  {Gil-White}, \citenamefont {Gurven}, \citenamefont {Marlowe}, \citenamefont
  {Patton},\ and\ \citenamefont {Tracer}}]{HBB_2005_BBSc}%
  \BibitemOpen
  \bibfield  {author} {\bibinfo {author} {\bibfnamefont {J.}~\bibnamefont
  {Henrich}}, \bibinfo {author} {\bibfnamefont {R.}~\bibnamefont {Boyd}},
  \bibinfo {author} {\bibfnamefont {S.}~\bibnamefont {Bowles}}, \bibinfo
  {author} {\bibfnamefont {C.}~\bibnamefont {Camerer}}, \bibinfo {author}
  {\bibfnamefont {E.}~\bibnamefont {Fehr}}, \bibinfo {author} {\bibfnamefont
  {H.}~\bibnamefont {Gintis}}, \bibinfo {author} {\bibfnamefont
  {R.}~\bibnamefont {McElreath}}, \bibinfo {author} {\bibfnamefont
  {M.}~\bibnamefont {Alvard}}, \bibinfo {author} {\bibfnamefont
  {A.}~\bibnamefont {Barr}}, \bibinfo {author} {\bibfnamefont {J.}~\bibnamefont
  {Ensminger}}, \bibinfo {author} {\bibfnamefont {N.~S.}\ \bibnamefont
  {Henrich}}, \bibinfo {author} {\bibfnamefont {K.}~\bibnamefont {Hill}},
  \bibinfo {author} {\bibfnamefont {F.}~\bibnamefont {Gil-White}}, \bibinfo
  {author} {\bibfnamefont {M.}~\bibnamefont {Gurven}}, \bibinfo {author}
  {\bibfnamefont {F.~W.}\ \bibnamefont {Marlowe}}, \bibinfo {author}
  {\bibfnamefont {J.~Q.}\ \bibnamefont {Patton}},\ and\ \bibinfo {author}
  {\bibfnamefont {D.}~\bibnamefont {Tracer}},\ }\bibfield  {title} {\bibinfo
  {title} {“economic man” in cross-cultural perspective: Behavioral
  experiments in 15 small-scale societies},\ }\href
  {https://doi.org/10.1017/s0140525x05000142} {\bibfield  {journal} {\bibinfo
  {journal} {Behavioral and Brain Sciences}\ }\textbf {\bibinfo {volume}
  {28}},\ \bibinfo {pages} {795–815} (\bibinfo {year} {2005})}\BibitemShut
  {NoStop}%
\bibitem [{\citenamefont {Binmore}\ and\ \citenamefont
  {Samuelson}(1994)}]{KS_1994_JITE}%
  \BibitemOpen
  \bibfield  {author} {\bibinfo {author} {\bibfnamefont {K.}~\bibnamefont
  {Binmore}}\ and\ \bibinfo {author} {\bibfnamefont {L.}~\bibnamefont
  {Samuelson}},\ }\bibfield  {title} {\bibinfo {title} {An economist's
  perspective on the evolution of norms},\ }\href
  {http://www.jstor.org/stable/40753015} {\bibfield  {journal} {\bibinfo
  {journal} {Journal of Institutional and Theoretical Economics (JITE)}\
  }\textbf {\bibinfo {volume} {150}},\ \bibinfo {pages} {45} (\bibinfo {year}
  {1994})}\BibitemShut {NoStop}%
\bibitem [{\citenamefont {Croson}(1996)}]{C_1996_JEBO}%
  \BibitemOpen
  \bibfield  {author} {\bibinfo {author} {\bibfnamefont {R.~T.}\ \bibnamefont
  {Croson}},\ }\bibfield  {title} {\bibinfo {title} {Information in ultimatum
  games: An experimental study},\ }\href
  {https://doi.org/10.1016/s0167-2681(96)00857-8} {\bibfield  {journal}
  {\bibinfo  {journal} {Journal of Economic Behavior \& Organization}\ }\textbf
  {\bibinfo {volume} {30}},\ \bibinfo {pages} {197–212} (\bibinfo {year}
  {1996})}\BibitemShut {NoStop}%
\bibitem [{\citenamefont {Gaertig}\ \emph {et~al.}(2012)\citenamefont
  {Gaertig}, \citenamefont {Moser}, \citenamefont {Alguacil},\ and\
  \citenamefont {Ruz}}]{GMA_2012_FN}%
  \BibitemOpen
  \bibfield  {author} {\bibinfo {author} {\bibfnamefont {C.}~\bibnamefont
  {Gaertig}}, \bibinfo {author} {\bibfnamefont {A.}~\bibnamefont {Moser}},
  \bibinfo {author} {\bibfnamefont {S.}~\bibnamefont {Alguacil}},\ and\
  \bibinfo {author} {\bibfnamefont {M.}~\bibnamefont {Ruz}},\ }\bibfield
  {title} {\bibinfo {title} {Social information and economic decision-making in
  the ultimatum game},\ }\href {http://dx.doi.org/10.3389/fnins.2012.00103}
  {\bibfield  {journal} {\bibinfo  {journal} {Frontiers in Neuroscience}\
  }\textbf {\bibinfo {volume} {6}} (\bibinfo {year} {2012})}\BibitemShut
  {NoStop}%
\bibitem [{\citenamefont {G\"{u}th}\ and\ \citenamefont
  {Kocher}(2014)}]{GK_2014_JEBO}%
  \BibitemOpen
  \bibfield  {author} {\bibinfo {author} {\bibfnamefont {W.}~\bibnamefont
  {G\"{u}th}}\ and\ \bibinfo {author} {\bibfnamefont {M.~G.}\ \bibnamefont
  {Kocher}},\ }\bibfield  {title} {\bibinfo {title} {More than thirty years of
  ultimatum bargaining experiments: Motives, variations, and a survey of the
  recent literature},\ }\href {https://doi.org/10.1016/j.jebo.2014.06.006}
  {\bibfield  {journal} {\bibinfo  {journal} {Journal of Economic Behavior \&
  Organization}\ }\textbf {\bibinfo {volume} {108}},\ \bibinfo {pages}
  {396–409} (\bibinfo {year} {2014})}\BibitemShut {NoStop}%
\bibitem [{\citenamefont {Chiang}(2008)}]{C_2008_RS}%
  \BibitemOpen
  \bibfield  {author} {\bibinfo {author} {\bibfnamefont {Y.-S.}\ \bibnamefont
  {Chiang}},\ }\bibfield  {title} {\bibinfo {title} {A path toward fairness:
  Preferential association and the evolution of strategies in the ultimatum
  game},\ }\href {https://doi.org/10.1177/1043463108089544} {\bibfield
  {journal} {\bibinfo  {journal} {Rationality and Society}\ }\textbf {\bibinfo
  {volume} {20}},\ \bibinfo {pages} {173–201} (\bibinfo {year}
  {2008})}\BibitemShut {NoStop}%
\bibitem [{\citenamefont {Zhang}\ \emph {et~al.}(2023)\citenamefont {Zhang},
  \citenamefont {Yang}, \citenamefont {Chen}, \citenamefont {Bai},\ and\
  \citenamefont {Xie}}]{ZYC_2023_CSF}%
  \BibitemOpen
  \bibfield  {author} {\bibinfo {author} {\bibfnamefont {Y.}~\bibnamefont
  {Zhang}}, \bibinfo {author} {\bibfnamefont {S.}~\bibnamefont {Yang}},
  \bibinfo {author} {\bibfnamefont {X.}~\bibnamefont {Chen}}, \bibinfo {author}
  {\bibfnamefont {Y.}~\bibnamefont {Bai}},\ and\ \bibinfo {author}
  {\bibfnamefont {G.}~\bibnamefont {Xie}},\ }\bibfield  {title} {\bibinfo
  {title} {Reputation update of responders efficiently promotes the evolution
  of fairness in the ultimatum game},\ }\href
  {https://doi.org/10.1016/j.chaos.2023.113218} {\bibfield  {journal} {\bibinfo
   {journal} {Chaos, Solitons \& Fractals}\ }\textbf {\bibinfo {volume}
  {169}},\ \bibinfo {pages} {113218} (\bibinfo {year} {2023})}\BibitemShut
  {NoStop}%
\bibitem [{\citenamefont {Szolnoki}\ \emph
  {et~al.}(2012{\natexlab{a}})\citenamefont {Szolnoki}, \citenamefont {Perc},\
  and\ \citenamefont {Szabó}}]{SPS_2012_PRL}%
  \BibitemOpen
  \bibfield  {author} {\bibinfo {author} {\bibfnamefont {A.}~\bibnamefont
  {Szolnoki}}, \bibinfo {author} {\bibfnamefont {M.}~\bibnamefont {Perc}},\
  and\ \bibinfo {author} {\bibfnamefont {G.}~\bibnamefont {Szabó}},\
  }\bibfield  {title} {\bibinfo {title} {Defense mechanisms of empathetic
  players in the spatial ultimatum game},\ }\bibfield  {journal} {\bibinfo
  {journal} {Physical Review Letters}\ }\textbf {\bibinfo {volume} {109}},\
  \href {https://doi.org/10.1103/physrevlett.109.078701}
  {10.1103/physrevlett.109.078701} (\bibinfo {year}
  {2012}{\natexlab{a}})\BibitemShut {NoStop}%
\bibitem [{\citenamefont {Szolnoki}\ \emph
  {et~al.}(2012{\natexlab{b}})\citenamefont {Szolnoki}, \citenamefont {Perc},\
  and\ \citenamefont {Szabó}}]{SPS_2012_EPL}%
  \BibitemOpen
  \bibfield  {author} {\bibinfo {author} {\bibfnamefont {A.}~\bibnamefont
  {Szolnoki}}, \bibinfo {author} {\bibfnamefont {M.}~\bibnamefont {Perc}},\
  and\ \bibinfo {author} {\bibfnamefont {G.}~\bibnamefont {Szabó}},\
  }\bibfield  {title} {\bibinfo {title} {Accuracy in strategy imitations
  promotes the evolution of fairness in the spatial ultimatum game},\ }\href
  {https://doi.org/10.1209/0295-5075/100/28005} {\bibfield  {journal} {\bibinfo
   {journal} {EPL (Europhysics Letters)}\ }\textbf {\bibinfo {volume} {100}},\
  \bibinfo {pages} {28005} (\bibinfo {year} {2012}{\natexlab{b}})}\BibitemShut
  {NoStop}%
\bibitem [{\citenamefont {Deng}\ \emph {et~al.}(2011)\citenamefont {Deng},
  \citenamefont {Tang},\ and\ \citenamefont {Zhang}}]{DTZ_2011_PHYSA}%
  \BibitemOpen
  \bibfield  {author} {\bibinfo {author} {\bibfnamefont {L.}~\bibnamefont
  {Deng}}, \bibinfo {author} {\bibfnamefont {W.}~\bibnamefont {Tang}},\ and\
  \bibinfo {author} {\bibfnamefont {J.}~\bibnamefont {Zhang}},\ }\bibfield
  {title} {\bibinfo {title} {The coevolutionary ultimatum game on different
  network topologies},\ }\href {https://doi.org/10.1016/j.physa.2011.06.076}
  {\bibfield  {journal} {\bibinfo  {journal} {Physica A: Statistical Mechanics
  and its Applications}\ }\textbf {\bibinfo {volume} {390}},\ \bibinfo {pages}
  {4227–4235} (\bibinfo {year} {2011})}\BibitemShut {NoStop}%
\bibitem [{\citenamefont {Snellman}\ \emph {et~al.}(2021)\citenamefont
  {Snellman}, \citenamefont {Barrio},\ and\ \citenamefont
  {Kaski}}]{SBK_2021_PHYSA}%
  \BibitemOpen
  \bibfield  {author} {\bibinfo {author} {\bibfnamefont {J.~E.}\ \bibnamefont
  {Snellman}}, \bibinfo {author} {\bibfnamefont {R.~A.}\ \bibnamefont
  {Barrio}},\ and\ \bibinfo {author} {\bibfnamefont {K.~K.}\ \bibnamefont
  {Kaski}},\ }\bibfield  {title} {\bibinfo {title} {Social structure formation
  in a network of agents playing a hybrid of ultimatum and dictator games},\
  }\href {https://doi.org/10.1016/j.physa.2020.125257} {\bibfield  {journal}
  {\bibinfo  {journal} {Physica A: Statistical Mechanics and its Applications}\
  }\textbf {\bibinfo {volume} {561}},\ \bibinfo {pages} {125257} (\bibinfo
  {year} {2021})}\BibitemShut {NoStop}%
\bibitem [{\citenamefont {Page}\ \emph {et~al.}(2000)\citenamefont {Page},
  \citenamefont {Nowak},\ and\ \citenamefont {Sigmund}}]{PNS_2000_PRSB}%
  \BibitemOpen
  \bibfield  {author} {\bibinfo {author} {\bibfnamefont {K.~M.}\ \bibnamefont
  {Page}}, \bibinfo {author} {\bibfnamefont {M.~A.}\ \bibnamefont {Nowak}},\
  and\ \bibinfo {author} {\bibfnamefont {K.}~\bibnamefont {Sigmund}},\
  }\bibfield  {title} {\bibinfo {title} {The spatial ultimatum game},\ }\href
  {https://doi.org/10.1098/rspb.2000.1266} {\bibfield  {journal} {\bibinfo
  {journal} {Proceedings of the Royal Society of London. Series B: Biological
  Sciences}\ }\textbf {\bibinfo {volume} {267}},\ \bibinfo {pages}
  {2177–2182} (\bibinfo {year} {2000})}\BibitemShut {NoStop}%
\bibitem [{\citenamefont {McKenzie~Alexander}(2007)}]{M_2007_BOOK}%
  \BibitemOpen
  \bibfield  {author} {\bibinfo {author} {\bibfnamefont {J.}~\bibnamefont
  {McKenzie~Alexander}},\ }\href {https://doi.org/10.1017/cbo9780511550997}
  {\emph {\bibinfo {title} {The Structural Evolution of Morality}}}\ (\bibinfo
  {publisher} {Cambridge University Press},\ \bibinfo {year}
  {2007})\BibitemShut {NoStop}%
\bibitem [{\citenamefont {Killingback}\ and\ \citenamefont
  {Studer}(2001)}]{KS_2001_PRSB}%
  \BibitemOpen
  \bibfield  {author} {\bibinfo {author} {\bibfnamefont {T.}~\bibnamefont
  {Killingback}}\ and\ \bibinfo {author} {\bibfnamefont {E.}~\bibnamefont
  {Studer}},\ }\bibfield  {title} {\bibinfo {title} {Spatial ultimatum games,
  collaborations and the evolution of fairness},\ }\href
  {https://doi.org/10.1098/rspb.2001.1697} {\bibfield  {journal} {\bibinfo
  {journal} {Proceedings of the Royal Society of London. Series B: Biological
  Sciences}\ }\textbf {\bibinfo {volume} {268}},\ \bibinfo {pages}
  {1797–1801} (\bibinfo {year} {2001})}\BibitemShut {NoStop}%
\bibitem [{\citenamefont {Wu}\ \emph {et~al.}(2013)\citenamefont {Wu},
  \citenamefont {Fu}, \citenamefont {Zhang},\ and\ \citenamefont
  {Wang}}]{WFZW_2013_SR}%
  \BibitemOpen
  \bibfield  {author} {\bibinfo {author} {\bibfnamefont {T.}~\bibnamefont
  {Wu}}, \bibinfo {author} {\bibfnamefont {F.}~\bibnamefont {Fu}}, \bibinfo
  {author} {\bibfnamefont {Y.}~\bibnamefont {Zhang}},\ and\ \bibinfo {author}
  {\bibfnamefont {L.}~\bibnamefont {Wang}},\ }\bibfield  {title} {\bibinfo
  {title} {Adaptive role switching promotes fairness in networked ultimatum
  game},\ }\href {http://dx.doi.org/10.1038/srep01550} {\bibfield  {journal}
  {\bibinfo  {journal} {Scientific Reports}\ }\textbf {\bibinfo {volume} {3}}
  (\bibinfo {year} {2013})}\BibitemShut {NoStop}%
\bibitem [{\citenamefont {Wang}\ \emph {et~al.}(2014)\citenamefont {Wang},
  \citenamefont {Chen},\ and\ \citenamefont {Wang}}]{WCW_2014_SR}%
  \BibitemOpen
  \bibfield  {author} {\bibinfo {author} {\bibfnamefont {X.}~\bibnamefont
  {Wang}}, \bibinfo {author} {\bibfnamefont {X.}~\bibnamefont {Chen}},\ and\
  \bibinfo {author} {\bibfnamefont {L.}~\bibnamefont {Wang}},\ }\bibfield
  {title} {\bibinfo {title} {Random allocation of pies promotes the evolution
  of fairness in the ultimatum game},\ }\href
  {http://dx.doi.org/10.1038/srep04534} {\bibfield  {journal} {\bibinfo
  {journal} {Scientific Reports}\ }\textbf {\bibinfo {volume} {4}} (\bibinfo
  {year} {2014})}\BibitemShut {NoStop}%
\bibitem [{\citenamefont {Zhang}\ \emph {et~al.}(2018)\citenamefont {Zhang},
  \citenamefont {Chen}, \citenamefont {Liu},\ and\ \citenamefont
  {Sun}}]{ZCL_2018_AMC}%
  \BibitemOpen
  \bibfield  {author} {\bibinfo {author} {\bibfnamefont {Y.}~\bibnamefont
  {Zhang}}, \bibinfo {author} {\bibfnamefont {X.}~\bibnamefont {Chen}},
  \bibinfo {author} {\bibfnamefont {A.}~\bibnamefont {Liu}},\ and\ \bibinfo
  {author} {\bibfnamefont {C.}~\bibnamefont {Sun}},\ }\bibfield  {title}
  {\bibinfo {title} {The effect of the stake size on the evolution of
  fairness},\ }\href {https://doi.org/10.1016/j.amc.2017.11.013} {\bibfield
  {journal} {\bibinfo  {journal} {Applied Mathematics and Computation}\
  }\textbf {\bibinfo {volume} {321}},\ \bibinfo {pages} {641–653} (\bibinfo
  {year} {2018})}\BibitemShut {NoStop}%
\bibitem [{\citenamefont {Gale}\ \emph {et~al.}(1995)\citenamefont {Gale},
  \citenamefont {Binmore},\ and\ \citenamefont {Samuelson}}]{GBS_1995_GEB}%
  \BibitemOpen
  \bibfield  {author} {\bibinfo {author} {\bibfnamefont {J.}~\bibnamefont
  {Gale}}, \bibinfo {author} {\bibfnamefont {K.~G.}\ \bibnamefont {Binmore}},\
  and\ \bibinfo {author} {\bibfnamefont {L.}~\bibnamefont {Samuelson}},\
  }\bibfield  {title} {\bibinfo {title} {Learning to be imperfect: The
  ultimatum game},\ }\href {https://doi.org/10.1016/s0899-8256(05)80017-x}
  {\bibfield  {journal} {\bibinfo  {journal} {Games and Economic Behavior}\
  }\textbf {\bibinfo {volume} {8}},\ \bibinfo {pages} {56–90} (\bibinfo
  {year} {1995})}\BibitemShut {NoStop}%
\bibitem [{\citenamefont {Rand}\ \emph {et~al.}(2013)\citenamefont {Rand},
  \citenamefont {Tarnita}, \citenamefont {Ohtsuki},\ and\ \citenamefont
  {Nowak}}]{RTON_2013_PNAS}%
  \BibitemOpen
  \bibfield  {author} {\bibinfo {author} {\bibfnamefont {D.~G.}\ \bibnamefont
  {Rand}}, \bibinfo {author} {\bibfnamefont {C.~E.}\ \bibnamefont {Tarnita}},
  \bibinfo {author} {\bibfnamefont {H.}~\bibnamefont {Ohtsuki}},\ and\ \bibinfo
  {author} {\bibfnamefont {M.~A.}\ \bibnamefont {Nowak}},\ }\bibfield  {title}
  {\bibinfo {title} {Evolution of fairness in the one-shot anonymous ultimatum
  game},\ }\href {https://doi.org/10.1073/pnas.1214167110} {\bibfield
  {journal} {\bibinfo  {journal} {Proceedings of the National Academy of
  Sciences}\ }\textbf {\bibinfo {volume} {110}},\ \bibinfo {pages}
  {2581–2586} (\bibinfo {year} {2013})}\BibitemShut {NoStop}%
\bibitem [{\citenamefont {Page}(2002)}]{P_2002_BMB}%
  \BibitemOpen
  \bibfield  {author} {\bibinfo {author} {\bibfnamefont {K.}~\bibnamefont
  {Page}},\ }\bibfield  {title} {\bibinfo {title} {Empathy leads to fairness},\
  }\href {https://doi.org/10.1006/bulm.2002.0321} {\bibfield  {journal}
  {\bibinfo  {journal} {Bulletin of Mathematical Biology}\ }\textbf {\bibinfo
  {volume} {64}},\ \bibinfo {pages} {1101–1116} (\bibinfo {year}
  {2002})}\BibitemShut {NoStop}%
\bibitem [{\citenamefont {Tamarit}\ and\ \citenamefont
  {Sánchez}(2016)}]{TS_2016_PO}%
  \BibitemOpen
  \bibfield  {author} {\bibinfo {author} {\bibfnamefont {I.}~\bibnamefont
  {Tamarit}}\ and\ \bibinfo {author} {\bibfnamefont {A.}~\bibnamefont
  {Sánchez}},\ }\bibfield  {title} {\bibinfo {title} {Emotions and strategic
  behaviour: The case of the ultimatum game},\ }\href
  {https://doi.org/10.1371/journal.pone.0158733} {\bibfield  {journal}
  {\bibinfo  {journal} {PLOS ONE}\ }\textbf {\bibinfo {volume} {11}},\ \bibinfo
  {pages} {e0158733} (\bibinfo {year} {2016})}\BibitemShut {NoStop}%
\bibitem [{\citenamefont {Slembeck}(1999{\natexlab{a}})}]{S_1999_Experimental}%
  \BibitemOpen
  \bibfield  {author} {\bibinfo {author} {\bibfnamefont {T.}~\bibnamefont
  {Slembeck}},\ }\href {https://doi.org/None} {\emph {\bibinfo {title}
  {Reputations and Fairness in Bargaining - Experimental Evidence from a
  Repeated Ultimatum Game With Fixed Opponents}}},\ \bibinfo {type}
  {Experimental}\ \bibinfo {number} {9905002}\ (\bibinfo  {institution}
  {University Library of Munich, Germany},\ \bibinfo {year} {1999})\BibitemShut
  {NoStop}%
\bibitem [{\citenamefont {Axelrod}\ and\ \citenamefont
  {Hamilton}(1981{\natexlab{a}})}]{AH_1981_AAAS}%
  \BibitemOpen
  \bibfield  {author} {\bibinfo {author} {\bibfnamefont {R.}~\bibnamefont
  {Axelrod}}\ and\ \bibinfo {author} {\bibfnamefont {W.~D.}\ \bibnamefont
  {Hamilton}},\ }\bibfield  {title} {\bibinfo {title} {The evolution of
  cooperation},\ }\href {https://doi.org/10.1126/science.7466396} {\bibfield
  {journal} {\bibinfo  {journal} {Science}\ }\textbf {\bibinfo {volume}
  {211}},\ \bibinfo {pages} {1390–1396} (\bibinfo {year}
  {1981}{\natexlab{a}})}\BibitemShut {NoStop}%
\bibitem [{\citenamefont {Nowak}\ and\ \citenamefont
  {Sigmund}(1990)}]{NS_1990_SPRINGER}%
  \BibitemOpen
  \bibfield  {author} {\bibinfo {author} {\bibfnamefont {M.}~\bibnamefont
  {Nowak}}\ and\ \bibinfo {author} {\bibfnamefont {K.}~\bibnamefont
  {Sigmund}},\ }\bibfield  {title} {\bibinfo {title} {The evolution of
  stochastic strategies in the prisoner’s dilemma},\ }\href
  {https://doi.org/10.1007/bf00049570} {\bibfield  {journal} {\bibinfo
  {journal} {Acta Applicandae Mathematicae}\ }\textbf {\bibinfo {volume}
  {20}},\ \bibinfo {pages} {247–265} (\bibinfo {year} {1990})}\BibitemShut
  {NoStop}%
\bibitem [{\citenamefont {Axelrod}\ and\ \citenamefont
  {Hamilton}(1981{\natexlab{b}})}]{AH_1981_Sc}%
  \BibitemOpen
  \bibfield  {author} {\bibinfo {author} {\bibfnamefont {R.}~\bibnamefont
  {Axelrod}}\ and\ \bibinfo {author} {\bibfnamefont {W.~D.}\ \bibnamefont
  {Hamilton}},\ }\bibfield  {title} {\bibinfo {title} {The evolution of
  cooperation},\ }\href {https://doi.org/10.1126/science.7466396} {\bibfield
  {journal} {\bibinfo  {journal} {Science}\ }\textbf {\bibinfo {volume}
  {211}},\ \bibinfo {pages} {1390–1396} (\bibinfo {year}
  {1981}{\natexlab{b}})}\BibitemShut {NoStop}%
\bibitem [{\citenamefont {Sekiguchi}\ and\ \citenamefont
  {Ohtsuki}(2015)}]{SO_2015_DGA}%
  \BibitemOpen
  \bibfield  {author} {\bibinfo {author} {\bibfnamefont {T.}~\bibnamefont
  {Sekiguchi}}\ and\ \bibinfo {author} {\bibfnamefont {H.}~\bibnamefont
  {Ohtsuki}},\ }\bibfield  {title} {\bibinfo {title} {Fixation probabilities of
  strategies for bimatrix games in finite populations},\ }\href
  {https://doi.org/10.1007/s13235-015-0170-2} {\bibfield  {journal} {\bibinfo
  {journal} {Dynamic Games and Applications}\ }\textbf {\bibinfo {volume}
  {7}},\ \bibinfo {pages} {93–111} (\bibinfo {year} {2015})}\BibitemShut
  {NoStop}%
\bibitem [{\citenamefont {Traulsen}\ \emph {et~al.}(2006)\citenamefont
  {Traulsen}, \citenamefont {Nowak},\ and\ \citenamefont
  {Pacheco}}]{TNP_2006_PRE}%
  \BibitemOpen
  \bibfield  {author} {\bibinfo {author} {\bibfnamefont {A.}~\bibnamefont
  {Traulsen}}, \bibinfo {author} {\bibfnamefont {M.~A.}\ \bibnamefont
  {Nowak}},\ and\ \bibinfo {author} {\bibfnamefont {J.~M.}\ \bibnamefont
  {Pacheco}},\ }\bibfield  {title} {\bibinfo {title} {Stochastic dynamics of
  invasion and fixation},\ }\href
  {http://dx.doi.org/10.1103/PhysRevE.74.011909} {\bibfield  {journal}
  {\bibinfo  {journal} {Physical Review E}\ }\textbf {\bibinfo {volume} {74}}
  (\bibinfo {year} {2006})}\BibitemShut {NoStop}%
\bibitem [{\citenamefont {Nowak}(2006)}]{N_2006_BOOK}%
  \BibitemOpen
  \bibfield  {author} {\bibinfo {author} {\bibfnamefont {M.~A.}\ \bibnamefont
  {Nowak}},\ }\href {https://doi.org/10.2307/j.ctvjghw98} {\emph {\bibinfo
  {title} {Evolutionary Dynamics: Exploring the Equations of Life}}}\ (\bibinfo
   {publisher} {Harvard University Press, Cambridge.},\ \bibinfo {year}
  {2006})\BibitemShut {NoStop}%
\bibitem [{\citenamefont {Fudenberg}\ and\ \citenamefont
  {Imhof}(2006)}]{FI_2006_JET}%
  \BibitemOpen
  \bibfield  {author} {\bibinfo {author} {\bibfnamefont {D.}~\bibnamefont
  {Fudenberg}}\ and\ \bibinfo {author} {\bibfnamefont {L.~A.}\ \bibnamefont
  {Imhof}},\ }\bibfield  {title} {\bibinfo {title} {Imitation processes with
  small mutations},\ }\href {https://doi.org/10.1016/j.jet.2005.04.006}
  {\bibfield  {journal} {\bibinfo  {journal} {Journal of Economic Theory}\
  }\textbf {\bibinfo {volume} {131}},\ \bibinfo {pages} {251–262} (\bibinfo
  {year} {2006})}\BibitemShut {NoStop}%
\bibitem [{\citenamefont {Imhof}\ and\ \citenamefont
  {Nowak}(2009)}]{IN_2009_PRSB}%
  \BibitemOpen
  \bibfield  {author} {\bibinfo {author} {\bibfnamefont {L.~A.}\ \bibnamefont
  {Imhof}}\ and\ \bibinfo {author} {\bibfnamefont {M.~A.}\ \bibnamefont
  {Nowak}},\ }\bibfield  {title} {\bibinfo {title} {Stochastic evolutionary
  dynamics of direct reciprocity},\ }\href
  {https://doi.org/10.1098/rspb.2009.1171} {\bibfield  {journal} {\bibinfo
  {journal} {Proceedings of the Royal Society B: Biological Sciences}\ }\textbf
  {\bibinfo {volume} {277}},\ \bibinfo {pages} {463–468} (\bibinfo {year}
  {2009})}\BibitemShut {NoStop}%
\bibitem [{\citenamefont {Cressman}()}]{C_1992_BOOK}%
  \BibitemOpen
  \bibfield  {author} {\bibinfo {author} {\bibfnamefont {R.}~\bibnamefont
  {Cressman}},\ }\href {https://doi.org/10.1007/978-3-642-49981-4} {\emph
  {\bibinfo {title} {The Stability Concept of Evolutionary Game Theory: A
  Dynamic Approach}}}\ (\bibinfo  {publisher} {Springer Berlin
  Heidelberg})\BibitemShut {NoStop}%
\bibitem [{\citenamefont {Cressman}(1996)}]{C_1996_TPB}%
  \BibitemOpen
  \bibfield  {author} {\bibinfo {author} {\bibfnamefont {R.}~\bibnamefont
  {Cressman}},\ }\bibfield  {title} {\bibinfo {title} {Frequency-dependent
  stability for two-species interactions},\ }\href
  {https://doi.org/10.1006/tpbi.1996.0011} {\bibfield  {journal} {\bibinfo
  {journal} {Theoretical Population Biology}\ }\textbf {\bibinfo {volume}
  {49}},\ \bibinfo {pages} {189–210} (\bibinfo {year} {1996})}\BibitemShut
  {NoStop}%
\bibitem [{\citenamefont {Cressman}\ and\ \citenamefont
  {Tao}(2014)}]{CT_2014_PNAS}%
  \BibitemOpen
  \bibfield  {author} {\bibinfo {author} {\bibfnamefont {R.}~\bibnamefont
  {Cressman}}\ and\ \bibinfo {author} {\bibfnamefont {Y.}~\bibnamefont {Tao}},\
  }\bibfield  {title} {\bibinfo {title} {The replicator equation and other game
  dynamics},\ }\href {https://doi.org/10.1073/pnas.1400823111} {\bibfield
  {journal} {\bibinfo  {journal} {Proceedings of the National Academy of
  Sciences}\ }\textbf {\bibinfo {volume} {111}},\ \bibinfo {pages}
  {10810–10817} (\bibinfo {year} {2014})}\BibitemShut {NoStop}%
\bibitem [{\citenamefont {Kumar~Dubey}\ \emph {et~al.}(2026)\citenamefont
  {Kumar~Dubey}, \citenamefont {Chakraborty}, \citenamefont {Patra},\ and\
  \citenamefont {Chakraborty}}]{DCPC_2025_Chaos}%
  \BibitemOpen
  \bibfield  {author} {\bibinfo {author} {\bibfnamefont {V.}~\bibnamefont
  {Kumar~Dubey}}, \bibinfo {author} {\bibfnamefont {S.}~\bibnamefont
  {Chakraborty}}, \bibinfo {author} {\bibfnamefont {A.}~\bibnamefont {Patra}},\
  and\ \bibinfo {author} {\bibfnamefont {S.}~\bibnamefont {Chakraborty}},\
  }\bibfield  {title} {\bibinfo {title} {Evolutionarily stable strategy in
  asymmetric games: Dynamical and information-theoretical perspectives},\
  }\href {http://dx.doi.org/10.1063/5.0297611} {\bibfield  {journal} {\bibinfo
  {journal} {Chaos: An Interdisciplinary Journal of Nonlinear Science}\
  }\textbf {\bibinfo {volume} {36}} (\bibinfo {year} {2026})}\BibitemShut
  {NoStop}%
\bibitem [{\citenamefont {Taylor}\ and\ \citenamefont
  {Jonker}(1978)}]{TJ_1978_MB}%
  \BibitemOpen
  \bibfield  {author} {\bibinfo {author} {\bibfnamefont {P.~D.}\ \bibnamefont
  {Taylor}}\ and\ \bibinfo {author} {\bibfnamefont {L.~B.}\ \bibnamefont
  {Jonker}},\ }\bibfield  {title} {\bibinfo {title} {Evolutionary stable
  strategies and game dynamics},\ }\href
  {https://doi.org/10.1016/0025-5564(78)90077-9} {\bibfield  {journal}
  {\bibinfo  {journal} {Mathematical Biosciences}\ }\textbf {\bibinfo {volume}
  {40}},\ \bibinfo {pages} {145–156} (\bibinfo {year} {1978})}\BibitemShut
  {NoStop}%
\bibitem [{\citenamefont {Schuster}\ \emph {et~al.}(1981)\citenamefont
  {Schuster}, \citenamefont {Sigmund}, \citenamefont {Hofbauer}, \citenamefont
  {Gottlieb},\ and\ \citenamefont {Merz}}]{SSH_1981_BC}%
  \BibitemOpen
  \bibfield  {author} {\bibinfo {author} {\bibfnamefont {P.}~\bibnamefont
  {Schuster}}, \bibinfo {author} {\bibfnamefont {K.}~\bibnamefont {Sigmund}},
  \bibinfo {author} {\bibfnamefont {J.}~\bibnamefont {Hofbauer}}, \bibinfo
  {author} {\bibfnamefont {R.}~\bibnamefont {Gottlieb}},\ and\ \bibinfo
  {author} {\bibfnamefont {P.}~\bibnamefont {Merz}},\ }\bibfield  {title}
  {\bibinfo {title} {Selfregulation of behaviour in animal societies: Iii.
  games between two populations with selfinteraction},\ }\href
  {https://doi.org/10.1007/bf00326677} {\bibfield  {journal} {\bibinfo
  {journal} {Biological Cybernetics}\ }\textbf {\bibinfo {volume} {40}},\
  \bibinfo {pages} {17–25} (\bibinfo {year} {1981})}\BibitemShut {NoStop}%
\bibitem [{\citenamefont {Veller}\ and\ \citenamefont
  {Hayward}(2016)}]{VH_2016_JET}%
  \BibitemOpen
  \bibfield  {author} {\bibinfo {author} {\bibfnamefont {C.}~\bibnamefont
  {Veller}}\ and\ \bibinfo {author} {\bibfnamefont {L.~K.}\ \bibnamefont
  {Hayward}},\ }\bibfield  {title} {\bibinfo {title} {Finite-population
  evolution with rare mutations in asymmetric games},\ }\href
  {https://doi.org/10.1016/j.jet.2015.12.005} {\bibfield  {journal} {\bibinfo
  {journal} {Journal of Economic Theory}\ }\textbf {\bibinfo {volume} {162}},\
  \bibinfo {pages} {93–113} (\bibinfo {year} {2016})}\BibitemShut {NoStop}%
\bibitem [{\citenamefont {Anderson}(1972)}]{A_1972_Sc}%
  \BibitemOpen
  \bibfield  {author} {\bibinfo {author} {\bibfnamefont {P.~W.}\ \bibnamefont
  {Anderson}},\ }\bibfield  {title} {\bibinfo {title} {More is different:
  Broken symmetry and the nature of the hierarchical structure of science.},\
  }\href {https://doi.org/10.1126/science.177.4047.393} {\bibfield  {journal}
  {\bibinfo  {journal} {Science}\ }\textbf {\bibinfo {volume} {177}},\ \bibinfo
  {pages} {393–396} (\bibinfo {year} {1972})}\BibitemShut {NoStop}%
\bibitem [{\citenamefont {Castellano}\ \emph {et~al.}(2009)\citenamefont
  {Castellano}, \citenamefont {Fortunato},\ and\ \citenamefont
  {Loreto}}]{CFL_2009_RMP}%
  \BibitemOpen
  \bibfield  {author} {\bibinfo {author} {\bibfnamefont {C.}~\bibnamefont
  {Castellano}}, \bibinfo {author} {\bibfnamefont {S.}~\bibnamefont
  {Fortunato}},\ and\ \bibinfo {author} {\bibfnamefont {V.}~\bibnamefont
  {Loreto}},\ }\bibfield  {title} {\bibinfo {title} {Statistical physics of
  social dynamics},\ }\href {https://doi.org/10.1103/revmodphys.81.591}
  {\bibfield  {journal} {\bibinfo  {journal} {Reviews of Modern Physics}\
  }\textbf {\bibinfo {volume} {81}},\ \bibinfo {pages} {591–646} (\bibinfo
  {year} {2009})}\BibitemShut {NoStop}%
\bibitem [{\citenamefont {Nowak}\ \emph {et~al.}(2004)\citenamefont {Nowak},
  \citenamefont {Sasaki}, \citenamefont {Taylor},\ and\ \citenamefont
  {Fudenberg}}]{NST_2004_NAT}%
  \BibitemOpen
  \bibfield  {author} {\bibinfo {author} {\bibfnamefont {M.~A.}\ \bibnamefont
  {Nowak}}, \bibinfo {author} {\bibfnamefont {A.}~\bibnamefont {Sasaki}},
  \bibinfo {author} {\bibfnamefont {C.}~\bibnamefont {Taylor}},\ and\ \bibinfo
  {author} {\bibfnamefont {D.}~\bibnamefont {Fudenberg}},\ }\bibfield  {title}
  {\bibinfo {title} {Emergence of cooperation and evolutionary stability in
  finite populations},\ }\href {https://doi.org/10.1038/nature02414} {\bibfield
   {journal} {\bibinfo  {journal} {Nature}\ }\textbf {\bibinfo {volume}
  {428}},\ \bibinfo {pages} {646–650} (\bibinfo {year} {2004})}\BibitemShut
  {NoStop}%
\bibitem [{\citenamefont {Almeida}\ \emph {et~al.}(2014)\citenamefont
  {Almeida}, \citenamefont {da~Silva},\ and\ \citenamefont
  {Martinez}}]{ASM_2014_PHYSA}%
  \BibitemOpen
  \bibfield  {author} {\bibinfo {author} {\bibfnamefont {E.}~\bibnamefont
  {Almeida}}, \bibinfo {author} {\bibfnamefont {R.}~\bibnamefont {da~Silva}},\
  and\ \bibinfo {author} {\bibfnamefont {A.~S.}\ \bibnamefont {Martinez}},\
  }\bibfield  {title} {\bibinfo {title} {Strategies generalization and payoff
  fluctuation optimization in the iterated ultimatum game},\ }\href
  {https://doi.org/10.1016/j.physa.2014.06.032} {\bibfield  {journal} {\bibinfo
   {journal} {Physica A: Statistical Mechanics and its Applications}\ }\textbf
  {\bibinfo {volume} {412}},\ \bibinfo {pages} {54–65} (\bibinfo {year}
  {2014})}\BibitemShut {NoStop}%
\bibitem [{\citenamefont {Slembeck}(1999{\natexlab{b}})}]{Slembeck1999}%
  \BibitemOpen
  \bibfield  {author} {\bibinfo {author} {\bibfnamefont {T.}~\bibnamefont
  {Slembeck}},\ }\href {https://ideas.repec.org/p/wpa/wuwpex/9905002.html}
  {\emph {\bibinfo {title} {Reputations and fairness in bargaining:
  {Experimental} evidence from a repeated ultimatum game with fixed
  opponents}}},\ \bibinfo {type} {Tech. Rep.}\ \bibinfo {number} {9905002}\
  (\bibinfo  {institution} {University Library of Munich, Germany},\ \bibinfo
  {year} {1999})\BibitemShut {NoStop}%
\bibitem [{\citenamefont {Proctor}\ \emph {et~al.}(2013)\citenamefont
  {Proctor}, \citenamefont {Williamson}, \citenamefont {de~Waal},\ and\
  \citenamefont {Brosnan}}]{Proctor2013}%
  \BibitemOpen
  \bibfield  {author} {\bibinfo {author} {\bibfnamefont {D.}~\bibnamefont
  {Proctor}}, \bibinfo {author} {\bibfnamefont {R.~A.}\ \bibnamefont
  {Williamson}}, \bibinfo {author} {\bibfnamefont {F.~B.~M.}\ \bibnamefont
  {de~Waal}},\ and\ \bibinfo {author} {\bibfnamefont {S.~F.}\ \bibnamefont
  {Brosnan}},\ }\bibfield  {title} {\bibinfo {title} {Chimpanzees play the
  ultimatum game},\ }\href {https://doi.org/10.1073/pnas.1220806110} {\bibfield
   {journal} {\bibinfo  {journal} {Proc. Natl. Acad. Sci. U.S.A.}\ }\textbf
  {\bibinfo {volume} {110}},\ \bibinfo {pages} {2070} (\bibinfo {year}
  {2013})}\BibitemShut {NoStop}%
\bibitem [{\citenamefont {Nowak}\ and\ \citenamefont
  {Sigmund}(1994)}]{NS_1994_JTB}%
  \BibitemOpen
  \bibfield  {author} {\bibinfo {author} {\bibfnamefont {M.~A.}\ \bibnamefont
  {Nowak}}\ and\ \bibinfo {author} {\bibfnamefont {K.}~\bibnamefont
  {Sigmund}},\ }\bibfield  {title} {\bibinfo {title} {The alternating
  prisoner’s dilemma},\ }\href {https://doi.org/10.1006/jtbi.1994.1101}
  {\bibfield  {journal} {\bibinfo  {journal} {Journal of Theoretical Biology}\
  }\textbf {\bibinfo {volume} {168}},\ \bibinfo {pages} {219–226} (\bibinfo
  {year} {1994})}\BibitemShut {NoStop}%
\bibitem [{\citenamefont {Park}\ \emph {et~al.}(2022)\citenamefont {Park},
  \citenamefont {Nowak},\ and\ \citenamefont {Hilbe}}]{PNH_2022_NATC}%
  \BibitemOpen
  \bibfield  {author} {\bibinfo {author} {\bibfnamefont {P.~S.}\ \bibnamefont
  {Park}}, \bibinfo {author} {\bibfnamefont {M.~A.}\ \bibnamefont {Nowak}},\
  and\ \bibinfo {author} {\bibfnamefont {C.}~\bibnamefont {Hilbe}},\ }\bibfield
   {title} {\bibinfo {title} {Cooperation in alternating interactions with
  memory constraints},\ }\bibfield  {journal} {\bibinfo  {journal} {Nature
  Communications}\ }\textbf {\bibinfo {volume} {13}},\ \href
  {https://doi.org/10.1038/s41467-022-28336-2} {10.1038/s41467-022-28336-2}
  (\bibinfo {year} {2022})\BibitemShut {NoStop}%
\bibitem [{\citenamefont {Press}\ and\ \citenamefont
  {Dyson}(2012)}]{PD_2012_PNAS}%
  \BibitemOpen
  \bibfield  {author} {\bibinfo {author} {\bibfnamefont {W.~H.}\ \bibnamefont
  {Press}}\ and\ \bibinfo {author} {\bibfnamefont {F.~J.}\ \bibnamefont
  {Dyson}},\ }\bibfield  {title} {\bibinfo {title} {Iterated prisoner’s
  dilemma contains strategies that dominate any evolutionary opponent},\ }\href
  {https://doi.org/10.1073/pnas.1206569109} {\bibfield  {journal} {\bibinfo
  {journal} {Proceedings of the National Academy of Sciences}\ }\textbf
  {\bibinfo {volume} {109}},\ \bibinfo {pages} {10409–10413} (\bibinfo {year}
  {2012})}\BibitemShut {NoStop}%
\bibitem [{\citenamefont {LaPorte}\ \emph {et~al.}(2023)\citenamefont
  {LaPorte}, \citenamefont {Hilbe},\ and\ \citenamefont
  {Nowak}}]{LHN_2023_PLoS}%
  \BibitemOpen
  \bibfield  {author} {\bibinfo {author} {\bibfnamefont {P.}~\bibnamefont
  {LaPorte}}, \bibinfo {author} {\bibfnamefont {C.}~\bibnamefont {Hilbe}},\
  and\ \bibinfo {author} {\bibfnamefont {M.~A.}\ \bibnamefont {Nowak}},\
  }\bibfield  {title} {\bibinfo {title} {Adaptive dynamics of memory-one
  strategies in the repeated donation game},\ }\href
  {https://doi.org/10.1371/journal.pcbi.1010987} {\bibfield  {journal}
  {\bibinfo  {journal} {PLOS Computational Biology}\ }\textbf {\bibinfo
  {volume} {19}},\ \bibinfo {pages} {e1010987} (\bibinfo {year}
  {2023})}\BibitemShut {NoStop}%
\bibitem [{\citenamefont {Chen}\ \emph {et~al.}(2015)\citenamefont {Chen},
  \citenamefont {Wu}, \citenamefont {Li}, \citenamefont {Wu},\ and\
  \citenamefont {Wang}}]{CWL_2015_EPL}%
  \BibitemOpen
  \bibfield  {author} {\bibinfo {author} {\bibfnamefont {W.}~\bibnamefont
  {Chen}}, \bibinfo {author} {\bibfnamefont {T.}~\bibnamefont {Wu}}, \bibinfo
  {author} {\bibfnamefont {Z.}~\bibnamefont {Li}}, \bibinfo {author}
  {\bibfnamefont {N.}~\bibnamefont {Wu}},\ and\ \bibinfo {author}
  {\bibfnamefont {L.}~\bibnamefont {Wang}},\ }\bibfield  {title} {\bibinfo
  {title} {Heterogenous allocation of chips promotes fairness in the ultimatum
  game},\ }\href {https://doi.org/10.1209/0295-5075/109/68006} {\bibfield
  {journal} {\bibinfo  {journal} {EPL (Europhysics Letters)}\ }\textbf
  {\bibinfo {volume} {109}},\ \bibinfo {pages} {68006} (\bibinfo {year}
  {2015})}\BibitemShut {NoStop}%
\bibitem [{\citenamefont {Roth}\ and\ \citenamefont {Erev}(1995)}]{Roth1995}%
  \BibitemOpen
  \bibfield  {author} {\bibinfo {author} {\bibfnamefont {A.~E.}\ \bibnamefont
  {Roth}}\ and\ \bibinfo {author} {\bibfnamefont {I.}~\bibnamefont {Erev}},\
  }\bibfield  {title} {\bibinfo {title} {Learning in extensive-form games:
  Experimental data and simple dynamic models in the intermediate term},\
  }\href {https://doi.org/10.1016/s0899-8256(05)80020-x} {\bibfield  {journal}
  {\bibinfo  {journal} {Games and Economic Behavior}\ }\textbf {\bibinfo
  {volume} {8}},\ \bibinfo {pages} {164–212} (\bibinfo {year}
  {1995})}\BibitemShut {NoStop}%
\bibitem [{\citenamefont {Iosifescu}\ and\ \citenamefont
  {Theodorescu}(1965)}]{IT_1965_EBV}%
  \BibitemOpen
  \bibfield  {author} {\bibinfo {author} {\bibfnamefont {M.}~\bibnamefont
  {Iosifescu}}\ and\ \bibinfo {author} {\bibfnamefont {R.}~\bibnamefont
  {Theodorescu}},\ }\bibfield  {title} {\bibinfo {title} {On bush-mosteller
  stochastic models for learning},\ }\href
  {https://doi.org/10.1016/0022-2496(65)90025-8} {\bibfield  {journal}
  {\bibinfo  {journal} {Journal of Mathematical Psychology}\ }\textbf {\bibinfo
  {volume} {2}},\ \bibinfo {pages} {196–203} (\bibinfo {year}
  {1965})}\BibitemShut {NoStop}%
\bibitem [{\citenamefont {Camerer}\ and\ \citenamefont
  {Hua~Ho}(1999)}]{CH_1999_E}%
  \BibitemOpen
  \bibfield  {author} {\bibinfo {author} {\bibfnamefont {C.}~\bibnamefont
  {Camerer}}\ and\ \bibinfo {author} {\bibfnamefont {T.}~\bibnamefont
  {Hua~Ho}},\ }\bibfield  {title} {\bibinfo {title} {Experience-weighted
  attraction learning in normal form games},\ }\href
  {https://doi.org/10.1111/1468-0262.00054} {\bibfield  {journal} {\bibinfo
  {journal} {Econometrica}\ }\textbf {\bibinfo {volume} {67}},\ \bibinfo
  {pages} {827–874} (\bibinfo {year} {1999})}\BibitemShut {NoStop}%
\bibitem [{\citenamefont {Macy}\ and\ \citenamefont
  {Flache}(2002)}]{MF_2002_PNAS}%
  \BibitemOpen
  \bibfield  {author} {\bibinfo {author} {\bibfnamefont {M.~W.}\ \bibnamefont
  {Macy}}\ and\ \bibinfo {author} {\bibfnamefont {A.}~\bibnamefont {Flache}},\
  }\bibfield  {title} {\bibinfo {title} {Learning dynamics in social
  dilemmas},\ }\href {https://doi.org/10.1073/pnas.092080099} {\bibfield
  {journal} {\bibinfo  {journal} {Proceedings of the National Academy of
  Sciences}\ }\textbf {\bibinfo {volume} {99}},\ \bibinfo {pages} {7229–7236}
  (\bibinfo {year} {2002})}\BibitemShut {NoStop}%
\bibitem [{\citenamefont {Patra}\ \emph {et~al.}(2024)\citenamefont {Patra},
  \citenamefont {Sengupta}, \citenamefont {Paul},\ and\ \citenamefont
  {Chakraborty}}]{PSPC_2024_NJP}%
  \BibitemOpen
  \bibfield  {author} {\bibinfo {author} {\bibfnamefont {A.}~\bibnamefont
  {Patra}}, \bibinfo {author} {\bibfnamefont {S.}~\bibnamefont {Sengupta}},
  \bibinfo {author} {\bibfnamefont {A.}~\bibnamefont {Paul}},\ and\ \bibinfo
  {author} {\bibfnamefont {S.}~\bibnamefont {Chakraborty}},\ }\bibfield
  {title} {\bibinfo {title} {Inferring to cooperate: Evolutionary games with
  bayesian inferential strategies},\ }\href
  {https://doi.org/10.1088/1367-2630/ad4e5e} {\bibfield  {journal} {\bibinfo
  {journal} {New Journal of Physics}\ }\textbf {\bibinfo {volume} {26}},\
  \bibinfo {pages} {063003} (\bibinfo {year} {2024})}\BibitemShut {NoStop}%
\bibitem [{\citenamefont {Selten}(1975)}]{S_1975_IJGT}%
  \BibitemOpen
  \bibfield  {author} {\bibinfo {author} {\bibfnamefont {R.}~\bibnamefont
  {Selten}},\ }\bibfield  {title} {\bibinfo {title} {Reexamination of the
  perfectness concept for equilibrium points in extensive games},\ }\href
  {https://doi.org/10.1007/bf01766400} {\bibfield  {journal} {\bibinfo
  {journal} {International Journal of Game Theory}\ }\textbf {\bibinfo {volume}
  {4}},\ \bibinfo {pages} {25–55} (\bibinfo {year} {1975})}\BibitemShut
  {NoStop}%
\bibitem [{\citenamefont {Aliprantis}\ and\ \citenamefont
  {Topolyan}(2011)}]{AT_2011_EL}%
  \BibitemOpen
  \bibfield  {author} {\bibinfo {author} {\bibfnamefont {C.}~\bibnamefont
  {Aliprantis}}\ and\ \bibinfo {author} {\bibfnamefont {I.}~\bibnamefont
  {Topolyan}},\ }\bibfield  {title} {\bibinfo {title} {Trembling-hand myopia
  and trembling-hand perfection},\ }\href
  {https://doi.org/10.1016/j.econlet.2011.05.046} {\bibfield  {journal}
  {\bibinfo  {journal} {Economics Letters}\ }\textbf {\bibinfo {volume}
  {113}},\ \bibinfo {pages} {39–41} (\bibinfo {year} {2011})}\BibitemShut
  {NoStop}%
\bibitem [{\citenamefont {Ito}\ \emph {et~al.}(2017)\citenamefont {Ito},
  \citenamefont {McNamara}, \citenamefont {Yamauchi},\ and\ \citenamefont
  {Higginson}}]{IYH_2017_JTB}%
  \BibitemOpen
  \bibfield  {author} {\bibinfo {author} {\bibfnamefont {K.}~\bibnamefont
  {Ito}}, \bibinfo {author} {\bibfnamefont {J.~M.}\ \bibnamefont {McNamara}},
  \bibinfo {author} {\bibfnamefont {A.}~\bibnamefont {Yamauchi}},\ and\
  \bibinfo {author} {\bibfnamefont {A.~D.}\ \bibnamefont {Higginson}},\
  }\bibfield  {title} {\bibinfo {title} {The evolution of cooperation by
  negotiation in a noisy world},\ }\href {https://doi.org/10.1111/jeb.13030}
  {\bibfield  {journal} {\bibinfo  {journal} {Journal of Evolutionary Biology}\
  }\textbf {\bibinfo {volume} {30}},\ \bibinfo {pages} {603–615} (\bibinfo
  {year} {2017})}\BibitemShut {NoStop}%
\bibitem [{\citenamefont {Chakraborty}\ \emph {et~al.}(2024)\citenamefont
  {Chakraborty}, \citenamefont {Agarwal},\ and\ \citenamefont
  {Chakraborty}}]{CAC_2024_PRE}%
  \BibitemOpen
  \bibfield  {author} {\bibinfo {author} {\bibfnamefont {S.}~\bibnamefont
  {Chakraborty}}, \bibinfo {author} {\bibfnamefont {I.}~\bibnamefont
  {Agarwal}},\ and\ \bibinfo {author} {\bibfnamefont {S.}~\bibnamefont
  {Chakraborty}},\ }\bibfield  {title} {\bibinfo {title} {Replicator-mutator
  dynamics of the rock-paper-scissors game: Learning through mistakes},\ }\href
  {http://dx.doi.org/10.1103/PhysRevE.109.034404} {\bibfield  {journal}
  {\bibinfo  {journal} {Physical Review E}\ }\textbf {\bibinfo {volume} {109}}
  (\bibinfo {year} {2024})}\BibitemShut {NoStop}%
\bibitem [{\citenamefont {Hilbe}\ \emph {et~al.}(2018)\citenamefont {Hilbe},
  \citenamefont {{\v{S}}imsa}, \citenamefont {Chatterjee},\ and\ \citenamefont
  {Nowak}}]{HSC_2018_NAT}%
  \BibitemOpen
  \bibfield  {author} {\bibinfo {author} {\bibfnamefont {C.}~\bibnamefont
  {Hilbe}}, \bibinfo {author} {\bibfnamefont {{\v{S}}.}~\bibnamefont
  {{\v{S}}imsa}}, \bibinfo {author} {\bibfnamefont {K.}~\bibnamefont
  {Chatterjee}},\ and\ \bibinfo {author} {\bibfnamefont {M.~A.}\ \bibnamefont
  {Nowak}},\ }\bibfield  {title} {\bibinfo {title} {Evolution of cooperation in
  stochastic games},\ }\href {https://doi.org/10.1038/s41586-018-0277-x}
  {\bibfield  {journal} {\bibinfo  {journal} {Nature}\ }\textbf {\bibinfo
  {volume} {559}},\ \bibinfo {pages} {246–249} (\bibinfo {year}
  {2018})}\BibitemShut {NoStop}%
\end{thebibliography}%
\end{document}